\documentclass[11pt]{article}
\usepackage{amsmath,amssymb}
\usepackage{framed}

\begin{document}

\setcounter{page}{1}

\pagestyle{plain}

\begin{center}
\Large{\bf Some Aspects of the Tachyon Inflation with Superpotential in Confrontation with Planck2018 Data}\\
\small \vspace{1cm} {Narges
Rashidi}$^{}$\footnote{n.rashidi@umz.ac.ir} \\
\vspace{0.5cm} $^{}$ Department of Theoretical Physics, Faculty of
Science,
University of Mazandaran,\\
P. O. Box 47416-95447, Babolsar, IRAN\\
\end{center}

\begin{abstract}
We study the tachyon inflation in the presence of the superpotential
as an inflationary potential. We study the primordial perturbations
and their non-gaussian feature in the equilateral configuration. We
use the Planck2018 TT, TE, EE+lowE+lensing+BK14+BAO joint data at
$68\%$ CL and $95\%$ CL, to perform numerical analysis on the scalar
perturbations and seek for the observational viability of the
tachyon inflation with superpotential. We also check the
observational viability of the model by studying the tensor part of
the perturbations and comparing the results with Planck2018 TT, TE,
EE+lowE+lensing+BK14+BAO+ LIGO$\&$Virgo2016 joint data at $68\%$ CL
and $95\%$ CL. By studying the phase space of the model's
parameters, we predict the amplitude of the equilateral
non-gaussianity in this model. The reheating phase after inflation
is another issue that is explored in this paper. We show that, in
some ranges of the model's parameters, it is possible to have an
observationally viable tachyon model with superpotential.
\end{abstract}
\newpage

\section{Introduction}

Although the simple inflation model is the one driven by a canonical
single scalar
field~\cite{Gut81,Lin82,Alb82,Lin90,Lid00a,Lid97,Rio02,Lyt09,Mal03},
it is possible that a non-canonical scalar field be the field
responsible for driving the initial inflationary expansion of the
universe. These models, which contain the scalar field with
non-canonical kinetic energy, are usually named as $k$-inflation.
The interesting point about the $k$-inflation models is that these
models predict the scale-dependent and non-gaussian distributed
primordial perturbations. Tachyon inflation is one of the
$k$-inflation models with some interesting cosmological results
which has attained a lot of attention. In fact, by slow rolling of
the tachyon field, associated with the D-branes in string theory,
down its potential, the universe experiences the smooth evolution
from the accelerating phase of the expansion to the era dominated by
the non-relativistic fluid~\cite{Sen99,Sen02a,Sen02b,Gib02,Gar04}.
Note that, although there is no experimental evidence for the
existence of the tachyon field, there are also no indefeasible
arguments against it. On the other hand, since the inflation phase
occurred around Planck scale, it seems logical that the M-String
theory inspired fields could have an important role in that phase.
The tachyon field as the field running the inflation has some
interesting physical features and cosmological consequences that
expand our physical perspective.

In this regard, some authors have studied inflation in the models
with the tachyon field. For instance, the author of~\cite{Fei02} has
studied the power-law inflation with the tachyon field and found an
exact solution for the flat universe. Also, the authors
of~\cite{del09} have studied the intermediate inflation in the
presence of the tachyon field. The authors of~\cite{Noz13a} have
studied the inflation in both the minimally and non-minimally
coupled tachyon field and compared the results with observational
data. In~\cite{Bar15} the authors have studied the tachyon inflation
based on the large-$N$ formalism and by comparing the results with
the observational data, they have shown the viability of the model.
To study some more interesting works on the tachyon inflation
see~\cite{Noz14,Bou16,Rez17,Ras18,Ras20}.

An interesting aspect of studying the inflation model is the
perturbation and non-gaussian feature of its amplitude. The
prediction of the simple single field inflation with the canonical
scalar field is the almost scale-independent and gaussian
perturbations~\cite{Gut81,Lin82,Alb82,Lin90,Lid00a,Lid97,Rio02,Lyt09,Mal03}.
In fact, with the standard canonical scalar field or Higgs and with
potentials such as $\phi$, $\phi^{2}$, $\phi^{\frac{4}{3}}$, and so
on, the behavior of the tensor-to-scalar ratio versus the scalar
spectral index is not consistent with Planck2018 TT, TE, EE +lowE
+lensing +BK14+BAO datasets. In recent years some cosmologist have
shown interest in the inflation models predicting the scale
dependent and large non-gaussian distributed perturbations and have
presented some interesting works on this
issue~\cite{Mal03,Bar04,Che10,Fel11a,Fel11b,Noz15,Noz16,Noz17,Noz18b,Noj19,Noz19,Ras20a,Ras20}.
In this way, by considering the nonminimal or modified gravity
models, authors have tried to propose the models of inflation
leading to the observationally viable perturbation parameters. Also,
one may think about other scalar fields. The phantom field has
negative energy density and although the negative energy density has
been predicted in some realizations of the relativistic quantum
field theories, it is less favorable to consider it. However, with a
tachyon field, it is possible to have an inflation model with
observationally viable values of the scalar spectral index and
tensor-to-scalar ratio. Also, its equation of state parameter can be
$-1$ (corresponding to dark energy and early time inflationary
expanding universe) and $0$ (corresponding to dust or dark matter
components). In this way, the tachyonic field can run the
inflationary phase, and after generating a non-relativistic fluid
dominated phase, could be considered as the dark
matter~\cite{Gib02}. Another interesting property of the tachyon
field is that in the tachyon model, it is possible to have the
amplitude of the non-gaussianity of the order of $10^{1}$ and even
more (see section 4).

It should be noticed that the current observational data also have
released somewhat scale dependence of the primordial perturbation
and corresponding amplitudes of the non-gaussianity. In fact, the
scale dependence of the perturbation, presented by the scalar
spectral index, has been released as $n_{s}=0.9658\pm0.0038$ implied
by Planck2018 TT, TE, EE+lowE+lensing+BAO+BK14 data by considering
$\Lambda CDM+r+\frac{dn_{s}}{d\ln k}$ model~\cite{pl18a,pl18b}. The
mentioned data have also set a constraint on the tensor-to-scalar
ratio as $r<0.072$. Another important perturbation parameter in
studying the perturbation in the inflation models is the tensor
spectral index which shows the scale dependence of the tensor
perturbations. The observational constraint on this perturbation
parameter is as $-0.62<n_{T}<0.53$, obtained from Planck2018 TT, TE,
EE +lowE+lensing+BK14+BAO+LIGO and Virgo2016
data~\cite{pl18a,pl18b}. Planck2018 team has also released some
constraints on the amplitudes of the non-gaussianity. The released
constraint on the equilateral configurations of the non-gaussianity
is as $f_{NL}^{equil}=-26\pm 47$~\cite{pl19}. In this regard, by
considering the tachyon field as the field responsible for the
inflation, it is possible to check the observational viability of
this inflation model.

An important phase after the end of the inflation era is the
reheating process. Inflationary expansion of the universe continues
as long as $\mid \epsilon,\eta\mid \ll 1$ (corresponding to the flat
potential). As soon as $\epsilon$ or $\eta$ reaches unity, meaning
that the potential is no longer flat, the inflationary expansion
ends. Then, the inflaton reaches the minimum of the potential, where
it oscillates, loses its energy, and decays into the plasma of the
relativistic particles. By this process, the radiation component
dominates the energy density of the universe~\cite{Ab82,Do82,Al82}.
In fact, by considering the reheating phase, it is possible to
explain the cosmic origin of the production of the matter component
in the universe~\cite{Kof94,Kof97}. It has also been shown that, the
production of the cosmic relics (for instance, photons and
neutrinos) and matter-antimatter asymmetry observed in the universe,
can be explained by considering the reheating process after
inflation~\cite{Gi99,Wa18,Di04,Lo14}. During this era, the effective
equation of state parameter (for the massive inflaton) can be $-1$,
corresponding to the domination of the potential energy over the
kinetic energy, and $+1$, corresponding to the domination of the
kinetic energy over the potential energy. Other important parameters
in studying the reheating process are the e-folds number and the
temperature during this phase, which give some information about
this process. In Ref.~\cite{Am15} one can find a review on the
reheating.

In this paper, we study inflation with a special type of potential,
named superpotential. The idea of superpotential has been introduced
in~\cite{Beh00} (see also~\cite{De00,Csa00}). In
Refs~\cite{Noz13a,Dav02,Noz12,Noz13b}, the authors have used the
idea of superpotential to obtain some cosmological solutions. In our
previous work, we have used the superpotential as the potential
driving the inflation in a nonminimally coupled canonical scalar
field model. However, in this work, we study the tachyon inflation
with a superpotential given by
\begin{equation} \label{eq1}
V(\phi)=\frac{\kappa^{2}}{6}\Big(W(\phi)\Big)^2-\frac{1}{8}\left(\frac{dW(\phi)}{d\phi}\right)^{2}\,,
\end{equation}
where $W(\phi)$ is an arbitrary function of the scalar field. Note
that, we didn't involve any supersymmetry here. However, since the
form of the potential (\ref{eq1}) is similar to the one in the
supergravity theories, this potential has been named superpotential.
In this regard, the paper is organized as follows. In section 2, we
study the inflation in the tachyon model with superpotential and
obtain the main inflation parameters in terms of the superpotential.
In section 3, we present both the linear and non-linear
perturbations. We obtain the scalar and tensor spectral indices,
tensor-to-scalar ratio and the non-linearity parameter in the
equilateral configuration of the non-gaussianity. The observational
viability of the model in confrontation with the Planck2018 data set
is studied in section 4. In this way, we obtain some constraints on
the model's parameters. We study the reheating phase after inflation
in section 5 and show that, in some ranges of the model's
parameters, it is possible to have instantaneous reheating.

\section{Tachyon Inflation with Superpotential}
\subsection{Tachyon Model}

To study the tachyon inflation with superpotential, we start with
the following action
\begin{eqnarray}
\label{eq2} S=\int d^{4}x\,\sqrt{-g}
\Bigg[\frac{1}{2\kappa^{4}}R-V(\phi)\,\sqrt{1-2\,\lambda\,X}\Bigg]\,.
\end{eqnarray}
In the above action, $\kappa$ is the gravitational constant, $R$ is
the Ricci scalar, $V(\phi)$ is the potential of the tachyon field
which we assume to be the superpotential. The parameter $\lambda$ is
the warp factor which is a constant and also we define
$X=-\frac{1}{2}\,\partial_{\mu}\phi,\partial^{\mu}\phi$. To find the
Einstein's field equation, we vary action (\ref{eq2}) and get
\begin{eqnarray}
\label{eq3}
G_{\mu\nu}=\kappa^{2}\Bigg[-g_{\mu\nu}V(\phi)\sqrt{1-2\,\lambda
X}+\frac{\lambda\,
V(\phi)\partial_{\mu}\phi\,\partial_{\nu}\phi}{\sqrt{1-2\,\lambda
X}}\Bigg]\,,
\end{eqnarray}
which with the FRW metric as
\begin{equation}
\label{eq4} ds^{2}=-dt^{2}+a^{2}(t)\delta_{ij}dx^{i}dx^{j}\,,
\end{equation}
leads to the following Friedmann equation for the tachyon model
\begin{eqnarray}
\label{eq5}
3H^{2}=\frac{\kappa^{2}\,V(\phi)}{\sqrt{1-\lambda\,\dot{\phi}^{2}}}\,.
\end{eqnarray}
Note that, from now on, a dot shows the derivative of the parameter
with respect to the time and a prime shows the derivative of the
parameter with respect to the tachyon field. We also find the second
Friedmann equation as follows
\begin{eqnarray}
\label{eq6}
2\dot{H}+3H^{2}=\kappa^{2}\bigg[V(\phi)\,\sqrt{1-\lambda\,\dot{\phi}^{2}}\,\bigg]\,.
\end{eqnarray}
By varying action (\ref{eq2}) with respect to the tachyon field,
$\phi$, the equation of motion of this field is obtained as
\begin{equation}
\label{eq7}\frac{\lambda\,\ddot{\phi}}{1-\lambda\,\dot{\phi}^{2}}+3\,\lambda\,H\dot{\phi}
+\frac{V'(\phi)}{V(\phi)}=0\,.
\end{equation}
To proceed, considering that the scalar field is a function of time,
we assume the tachyon field as $\phi\equiv \phi(a)$. In this regard,
the Friedmann equations (\ref{eq5}) and (\ref{eq6}) take the
following forms
\begin{eqnarray}
\label{eq8}
3H^{2}=\frac{\kappa^{2}\,V(\phi)}{\sqrt{1-\lambda\,H^{2}\,\Big(a\frac{d\phi}{d
a}\Big)^{2}}}\,,
\end{eqnarray}
\begin{eqnarray}
\label{eq9} 2H\,H'\,a\,\frac{d\phi}{d
a}+3H^{2}=\kappa^{2}\bigg[V(\phi)\,\sqrt{1-\lambda\,H^{2}\,\Big(a\frac{d\phi}{d
a}\Big)^{2}}\,\bigg]\,.
\end{eqnarray}
Also, the equation of motion of the tachyon field becomes as
\begin{equation}
\label{eq10}\lambda\,\Bigg(\ddot{a}\,\frac{d\phi}{da}+\dot{a}^{2}\,\frac{d^{2}\phi}{d\phi^{2}}\Bigg)
+3\,\lambda\,H^2\,a\,\frac{d\phi}{da}\Bigg(1-\lambda\,H^{2}\,\Big(a\,\frac{d\phi}{da}\Big)^{2}\Bigg)
+\frac{V'(\phi)}{V(\phi)}\Bigg(1-\lambda\,H^{2}\,\Big(a\,\frac{d\phi}{da}\Big)^{2}\Bigg)=0\,.
\end{equation}
To study the inflation in this model, it is needed to calculate the
slow-roll parameters. The definition of the slow-roll parameters are
given by
\begin{eqnarray}
\label{eq11}\epsilon=-\frac{\dot{H}}{H^{2}}=-\frac{H}{H'}\,a\,\frac{d\phi}{da}\,,
\end{eqnarray}
\begin{eqnarray}
\label{eq12}\eta=\frac{1}{H}\frac{\dot{\epsilon}}{\epsilon}=\frac{\epsilon'}{\epsilon}\,a\,\frac{d\phi}{da}\,,
\end{eqnarray}
\begin{eqnarray}
\label{eq13}s=\frac{1}{H}\frac{\dot{c_{s}}}{c_{s}}=\frac{c'_{s}}{c_{s}}\,a\,\frac{d\phi}{da}\,.
\end{eqnarray}
In the last equation, the parameter $c_{s}$, defined as
$c_{s}^{2}\equiv\frac{P_{,X}}{\rho_{,X}}$, is the sound speed of the
perturbations where $_{,X}$ shows the derivative of the parameter
with respect to $X$. Considering that in the tachyon model we have
$P=-V(\phi)\sqrt{1-2\,\lambda\, X}$, we find the following
expression for the sound speed
\begin{eqnarray}
\label{eq14} c_{s}=\sqrt{1-\lambda\,H^{2}\,\Big(a\frac{d\phi}{d
a}\Big)^{2}}\,.
\end{eqnarray}
Another important parameter during inflation is the e-folds number,
which in this case is given by
\begin{eqnarray}
\label{eq15} N=\int
H\,dt=\int\frac{d\phi}{H\,a\,\frac{d\phi}{da}}\,.
\end{eqnarray}

\subsection{Superpotential}
In this subsection, we consider the potential defined in equation
(\ref{eq1}) and study the tachyon inflation. In this respect, to
equations (\ref{eq8})-(\ref{eq10}) be satisfied, we should have the
following expressions
\begin{eqnarray}
\label{eq16} \left(\frac{\dot{a}}{a}\right)=\kappa^{4}\,W^{2}\,,
\end{eqnarray}
\begin{eqnarray}
\label{eq17} a\,\frac{d\phi}{da}={\frac {\sqrt {\lambda\,W \left(
5168\,{\kappa}^{4}\,{W}^{4}+24\, {W}^{2}{{\it
W'}}^{2}{\kappa}^{2}-9\,{{\it W'}}^{4} \right) }}{72\,
\lambda\,{W}^{3}{\kappa}^{3}}} \,.
\end{eqnarray}
Now, by using the above equations, we can obtain equations
(\ref{eq11})-(\ref{eq13}) as follows
\begin{eqnarray}
\label{eq18} \epsilon=-{\frac { W'\, \sqrt {\lambda\,W \, \Big(
5168\,{\kappa}^ {4}\,W^{4}+24\, W^{2} \,W'^{2}{\kappa}^{2}-9\, W^{4}
\Big) }}{72\,{\kappa}^{3}\,W^{4}}} \,,
\end{eqnarray}
\begin{eqnarray}
\label{eq19} \eta= \frac {\sqrt { \left( 5168\,{\kappa}^{4}-9
\right)  W^{2}+24\,{\kappa}^{2}\, W'^{2}}}{W^{\frac{5}{2}} } \Bigg[
\sqrt {\lambda} \Big( 10336\, W^{3}{\kappa}^{4}+96\, W'^{2}\,W
\,{\kappa}^{2}- 18\, W^{3} \Big)
W''\nonumber\\
-120\,\sqrt { \lambda}{\kappa}^{2} W'^{4} -15504\, \left(
{\kappa}^{4}-{\frac{9}{5168}} \right) \sqrt {\lambda} \,W^{2}
\,W'^{2 } \Bigg]
\nonumber\\
\Bigg[ 3456\,{\kappa}^{5} W'^{3}+744192\, \left( {\kappa}^{4}-{
\frac{9}{5168}} \right) {\kappa}^{3} \,W^{2}\,W' \Bigg]^{-1} \,,
\end{eqnarray}

\begin{eqnarray}
\label{eq20} s= \frac{ \Big( 48\, W'\, {\kappa}^{2}\,W\,W''
-24\,{\kappa}^{2} \, W'^{3}+ W^{2} \left( 5168\,{\kappa}^{4}- 9
\right) W' \Big) { \lambda}^{\frac{5}{2}}\sqrt {24\,{\kappa}^{2}
W'^{2}+ W^{2} \left( 5168\,{\kappa}^{4}-9 \right) }}{
W^{\frac{5}{2}}\,\left( 3456\,{\kappa}^{5}{ \lambda}^{2}
\,W'^{2}+744192\,{\kappa}^{3} \left( {\kappa}^{4}-{\frac{9}{5168} }
\right) {\lambda}^{2} W^{2}- 746496\,{\kappa}^{5}W \right) }\,. \nonumber\\
\end{eqnarray}
We also find the following expression for the square of the sound
speed of the primordial perturbation in the tachyon model
\begin{eqnarray}
\label{eq21} c_{s}^{2}= {\frac {1}{72}\sqrt {-{\frac {{\lambda}^{2}
\left( 5168\,{\kappa}^{4}\,W^{4}+24\, W^{2}
\,W'^{2}\,{\kappa}^{2}-9\, W^{4} \right) }{
W^{3}\,{\kappa}^{2}}}+5184}} \,.
\end{eqnarray}
To use these equations and perform numerical analysis on the model,
we should adopt some specific function for $W(\phi)$. In this paper,
we choose the following form for $W(\phi)$
\begin{eqnarray}
\label{eq22}
W(\phi)=c\Bigg(\frac{e^{\alpha_{1}\,\phi}}{\alpha_{1}}-\frac{e^{\alpha_{2}\,\phi}}{\alpha_{2}}\Bigg)
\,,
\end{eqnarray}
where the parameters $c$, $\alpha_{1}$ and $\alpha_{2}$ are some
arbitrary constants. Note that, these parameters can be negative or
positive, however, in our model we consider the positive values of
these parameters. We also choose these parameters in the way that at
$\phi\rightarrow 0$, we have $H\rightarrow 1$. Therefore, the
following relation between the constant parameters is applied
\begin{eqnarray}
\label{eq23}
\alpha_{1}=\frac{c\,\kappa^{2}\,\alpha_{2}}{c\,\kappa^{2}+\alpha_{2}}\,.
\end{eqnarray}
By having the required relations, in the next section, we study the
perturbations in the tachyon model with superpotential to seek for
the observational viability of this model.

\section{Perturbations in the Tachyon Model with Superpotential}
\subsection{Linear Perturbations}
In this subsection, by considering the following perturbed ADM line
element
\begin{eqnarray}
\label{eq24} ds^{2}=
-(1+2{\cal{R}})dt^{2}+2a(t){\cal{M}}_{i}\,dt\,dx^{i}
+a^{2}(t)\left[(1-2{\Phi})\delta_{ij}+2{\Theta}_{ij}\right]dx^{i}dx^{j}\,,
\end{eqnarray}
we study the linear perturbation in this setup. In perturbed metric
(\ref{eq24}), the parameter ${\cal{M}}^{i}$ is defined as
${\cal{M}}^{i}=\delta^{ij}\partial_{j}{\cal{M}}+v^{i}$. The
condition $v^{i}_{,i}=0$ is satisfied by vector $v^{i}$. Also the
parameters ${\cal{R}}$ and ${\cal{M}}$ are 3-scalars~\cite{Muk92}.
In this perturbed metric, we have used the parameter $\Phi$ to
denote the spatial curvature perturbation and the parameter
${\Theta}_{ij}$ to show the spatial symmetric and traceless shear
3-tensor. By using the scalar part of the metric (\ref{eq24}) as
\begin{eqnarray}
\label{eq25}
ds^{2}=-(1+2{\cal{R}})dt^{2}+2a(t){\cal{M}}_{,i}\,dt\,dx^{i}
+a^{2}(t)(1-2{\Phi})\delta_{ij}dx^{i}dx^{j}\,,
\end{eqnarray}
written in the uniform-field gauge ($\delta\phi=0$) at the linear
level of the perturbations, we can explore the scalar perturbations
in this setup. To this end, and by using equation (\ref{eq25}), we
find the quadratic action as follows
\begin{equation}
\label{eq26} S_{2}=\int
dt\,d^{3}x\,a^{3}{\cal{Q}}\left[\bigg(\kappa^2\,W\,a\,\frac{d\Phi}{da}\bigg)^{2}-\frac{c_{s}^{2}}{a^{2}}(\partial
{\Phi})^{2}\right],
\end{equation}
which is obtained by expanding action (\ref{eq2}) up to the
second-order of the small perturbations. In the above quadratic
action, the parameter ${\cal{Q}}$ is defined as
\begin{eqnarray}
\label{eq27} {\cal{Q}}={\frac { V \, \lambda\,\dot{\phi}^{2}
}{2\,H^{2} \left(1- \lambda\,\dot{\phi}^{2} \right)^{\frac{3}{2}}
}}\hspace{9cm}\nonumber\\
=\frac { \Big( -\frac{1}{8}\,
W'^{2}+\frac{1}{6}\,{\kappa}^{2}\,{W}^{2} \Big) {\lambda}^ {2} \Big(
5168\,{\kappa}^{4} \,W^{ 4}+24\, W^{2}\, W'^{2}\,{\kappa}^{2}-9 \,
W^{4} \Big) }{10368\, \left( W \left( \phi \right)  \right)
^{7}{\kappa}^{10}} \nonumber\\
\times\Bigg[ -{ \frac {{\lambda}^{2} \Big( 5168\,{\kappa}^{4}
W^{4}+24\, W^{2}, W'^{2 }\,{\kappa}^{2}-9\, W^{4} \Big) }{5184\,
W^{5}\,{\kappa}^{6}}}+1 \Bigg]^{-\frac{3}{2}}\,.
\end{eqnarray}
Also, the sound speed squared is given by equation (\ref{eq21}). To
find more information about obtaining the higher-order actions,
see~\cite{Fel11a,Fel11b,Noz15}. One important parameter in studying
the scalar part of the perturbation is the scalar spectral index and
to obtain this parameter, we should use the following two-point
correlation function
\begin{equation}
\label{eq28} \langle
0|{\Phi}(0,\textbf{k}_{1}){\Phi}(0,\textbf{k}_{2})|0\rangle
=(2\pi)^{3}\delta^{(3)}(\textbf{k}_{1}+\textbf{k}_{2})\frac{2\pi^{2}}{k^{3}}{\cal{A}}_{s}\,,
\end{equation}
with
\begin{equation}
\label{eq29}
{\cal{A}}_{s}=\frac{H^{2}}{8\pi^{2}{\cal{Q}}c_{s}^{3}}\,,
\end{equation}
which is named the power spectrum. The power spectrum gives the
scalar spectral index as
\begin{equation}
\label{eq30} n_{s}-1=\frac{d \ln {\cal{A}}_{s}}{d \ln
k}\Bigg|_{c_{s}k=aH}\,.
\end{equation}
The following expression gives the scalar spectral index in terms of
the slow-roll parameters
\begin{equation}
\label{eq31} n_{s}=1-2\epsilon-\eta-s\,.
\end{equation}

The second-order action of the tensor mode is obtained as
\begin{eqnarray}
\label{eq32} S_{T}=\int dt\, d^{3}x\, \frac{a^{3}}{4\kappa^{2}}
\Bigg[\dot{\Theta}_{+}^{2}-\frac{1}{a^{2}}(\partial
{\Theta}_{+})^{2}+\dot{\Theta}_{\times}^{2}-\frac{1}{a^{2}}(\partial
{\Theta}_{\times})^{2}\Bigg]\,,
\end{eqnarray}
where we have used two polarization tensors
$\vartheta_{ij}^{(+,\times)}$, to rewrite the 3-tensor
${\Theta}_{ij}$ as
${\Theta}_{ij}={\Theta}_{+}\vartheta_{ij}^{+}+{\Theta}_{\times}\vartheta_{ij}^{\times}$.
The amplitude of the tensor perturbations is obtained by following
the method used in the scalar part as follows
\begin{equation}
\label{eq33} {\cal{A}}_{T}=\frac{2\kappa^{2}H^{2}}{\pi^{2}}\,,
\end{equation}
giving the following tensor spectral index in this model
\begin{equation}
\label{eq34} n_{T}=\frac{d \ln {\cal{A}}_{T}}{d \ln k}=-2\epsilon\,.
\end{equation}

The ratio between the amplitude of the tensor spectral index and the
amplitude of the scalar spectral index gives another important
parameter as
\begin{equation}
\label{eq35}
r=\frac{{\cal{A}}_{T}}{{\cal{A}}_{s}}=16\,c_{s}\,\epsilon\,,
\end{equation}
named the tensor-to-scalar ratio. Note that, in equations
(\ref{eq31}), (\ref{eq33}) and (\ref{eq35}), the slow-roll
parameters are given by equations (\ref{eq18})-(\ref{eq20}).

The perturbations parameters obtained in this subsection, are
important in studying the linear perturbations. To find more
information about the perturbations, it is useful to study the
non-linear level of the perturbation. In this regard, in the next
subsection, we explore the non-linear perturbations to get some
information about the non-gaussian feature of the primordial
perturbations in the tachyon model with superpotential.

\subsection{Non-linear Perturbations}

In this subsection, we go to the non-linear level of the
perturbations to study the non-gaussianity in this model. By
introducing the new parameter $\Upsilon$ which satisfies the
following relations
\begin{eqnarray}
\label{eq36}
{\cal{M}}=\frac{{\Phi}}{\kappa^{2}\,W}+\kappa^{2}a^{2}\Upsilon\,,
\end{eqnarray}
and
\begin{equation}
\label{eq37}
\partial^{2}\Upsilon={\cal{Q}}\,\kappa^{2}\,W\,a\frac{d\Phi}{da}\,,
\end{equation}
and expanding action (\ref{eq2}) up to the third order in the
perturbations, we get
\begin{eqnarray}
\label{eq38} S_{3}=\int dt\, d^{3}x\,\Bigg\{
\Bigg[\frac{3a^{3}}{\kappa^{2}c_{s}^{2}}\,
\Bigg(1-\frac{1}{c_{s}^{2}}\Bigg) \epsilon
\Bigg]{\Phi}\,\bigg(\kappa^{2}\,W\,a\frac{d\Phi}{da}\bigg)^{2}
+\Bigg[\frac{a}{\kappa^{2}} \Bigg(\frac{1}{c_{s}^{2}}-1\Bigg)
\epsilon
\Bigg]{\Phi}\,(\partial{\Phi})^{2}\nonumber\\
+\Bigg[\frac{a^{3}}{\kappa^{2}}\, \Bigg(\frac{1}{c_{s}^{2}\,H}\Bigg)
\Bigg(\frac{1}{c_{s}^{2}}-1\Bigg)\epsilon\Bigg]
\bigg(\kappa^{2}\,W\,a\frac{d\Phi}{da}\bigg)^{3}-\Bigg[a^{3}\,\frac{2}{c_{s}^{2}}\epsilon\,\bigg(\kappa^{2}\,W\,a\frac{d\Phi}{da}\bigg)
(\partial_{i}{\Phi})(\partial_{i}\Upsilon)\Bigg]\Bigg\}\,,
\end{eqnarray}
which is dubbed the cubic action and is written up to the leading
order in the slow-roll parameters of the model. Now, we use the
three-point correlation function for the spatial curvature
perturbation in the interaction picture as~\cite{Mal03,Che08}
\begin{eqnarray}
\label{eq39} \langle
{\Phi}(\textbf{k}_{1})\,{\Phi}(\textbf{k}_{2})\,{\Phi}(\textbf{k}_{3})\rangle
=(2\pi)^{3}\delta^{3}(\textbf{k}_{1}+\textbf{k}_{2}+\textbf{k}_{3}){\cal{B}}_{\Phi}(\textbf{k}_{1},\textbf{k}_{2},\textbf{k}_{3})\,,
\end{eqnarray}
with following definition for ${\cal{B}}_{\Phi}$
\begin{equation}
\label{eq40}
{\cal{B}}_{\Phi}(\textbf{k}_{1},\textbf{k}_{2},\textbf{k}_{3})=\frac{(2\pi)^{4}{\cal{A}}_{s}^{2}}{\prod_{i=1}^{3}
k_{i}^{3}}\,
{\cal{G}}_{\Phi}(\textbf{k}_{1},\textbf{k}_{2},\textbf{k}_{3})\,.
\end{equation}
Also, the parameter ${\cal{G}}_{\Phi}$ in equation (\ref{eq40}) is
given by the following expression
\begin{eqnarray}
\label{eq41} {\cal{G}}_{\Phi}=\Bigg(1-\frac{1}{c_{s}^{2}}\Bigg)
\Bigg[\frac{3}{4}\Bigg(\frac{2\sum_{i>j}k_{i}^{2}\,k_{j}^{2}}{k_{1}+k_{2}+k_{3}}-\frac{\sum_{i\neq
j}k_{i}^{2}\,k_{j}^{3}}{(k_{1}+k_{2}+k_{3})^{2}}\Bigg)-\frac{3}{2}\Bigg(\frac{\left(k_{1}\,k_{2}\,k_{3}\right)^{2}}
{(k_{1}+k_{2}+k_{3})^{3}}\Bigg)\nonumber\\
-\frac{1}{4}\Bigg(
\frac{2\sum_{i>j}k_{i}^{2}\,k_{j}^{2}}{k_{1}+k_{2}+k_{3}}-\frac{\sum_{i\neq
j}k_{i}^{2}\,k_{j}^{3}}{(k_{1}+k_{2}+k_{3})^{2}}+\frac{1}{2}\sum_{i}k_{i}^{3}\Bigg)
\Bigg]\,.
\end{eqnarray}

In studying the non-gaussian feature of the primordial
perturbations, it is useful to introduce the so-called
``non-linearity parameter''. This parameter is defined by using the
parameter ${\cal{G}}_{\Phi}$ as follows
\begin{equation}
\label{eq42}
f_{NL}=\frac{10}{3}\frac{{\cal{G}}_{\Phi}}{\sum_{i=1}^{3}k_{i}^{3}}\,.
\end{equation}
As it is clear from equation (\ref{eq42}), the non-linearity
parameter depends on the values of different momenta $k_{1}$,
$k_{2}$ and $k_{3}$, which lead to different shapes of the
primordial non-gaussianity with the different maximal signal of
their amplitudes. Considering that the signal of the amplitude of
the non-gaussianity in the $k$-inflation and higher-order derivative
models becomes maximal at the equilateral configuration, in this
paper we consider
$k_{1}=k_{2}=k_{3}$~\cite{Che07,Bab04b,Fel13a,Bau12}. By adopting
the equilateral configuration, we find
\begin{equation}
\label{eq43}
{\cal{G}}_{\Phi}^{equil}=\frac{17}{72}k^3\left(1-\frac{1}{c_{s}^{2}}\right)\,,
\end{equation}
and therefore
\begin{equation}
\label{eq44}
f^{equil}_{NL}=\frac{85}{324}\left(1-\frac{1}{c_{s}^{2}}\right)\,,
\end{equation}
where $c_{s}^{2}$ is given by equation (\ref{eq21}). Now that we
have the important parameters in both linear and non-linear levels,
we can seek for the observational viability of the tachyon inflation
with superpotential. In this regard, in the next section, we present
the numerical analysis of this setup.

\begin{figure}
\begin{center}\includegraphics{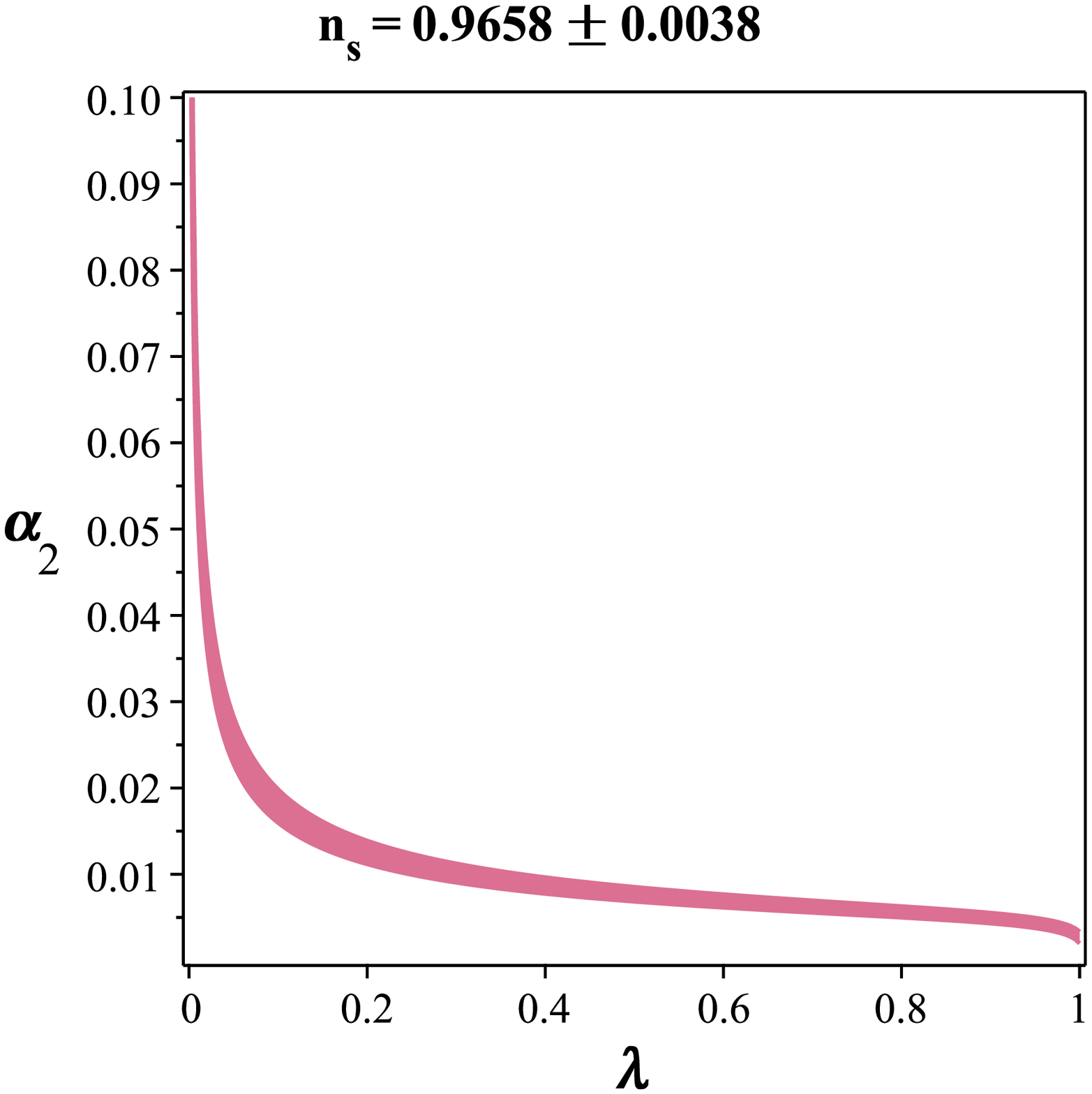}\includegraphics{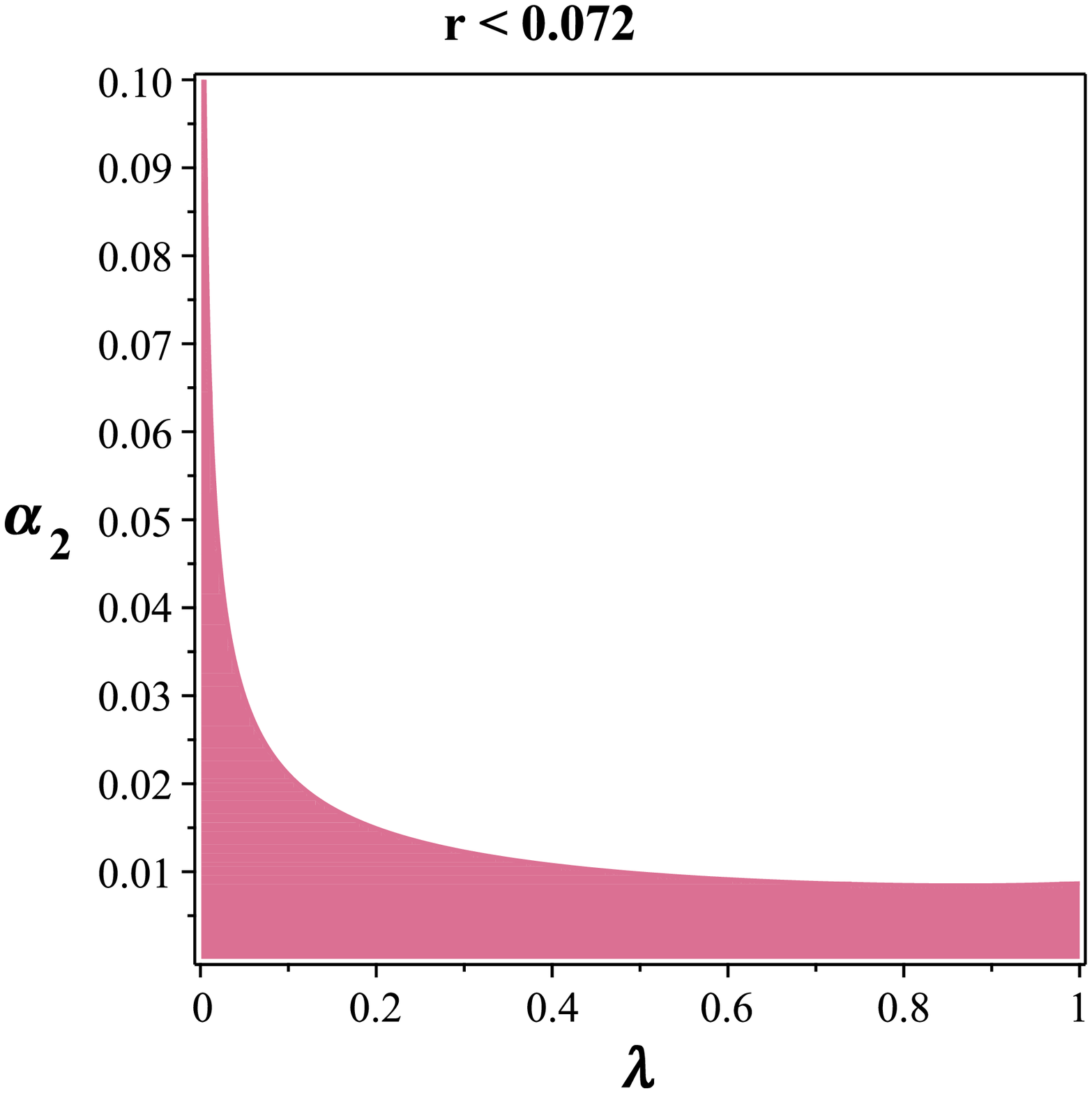}\vspace{6cm}
\includegraphics{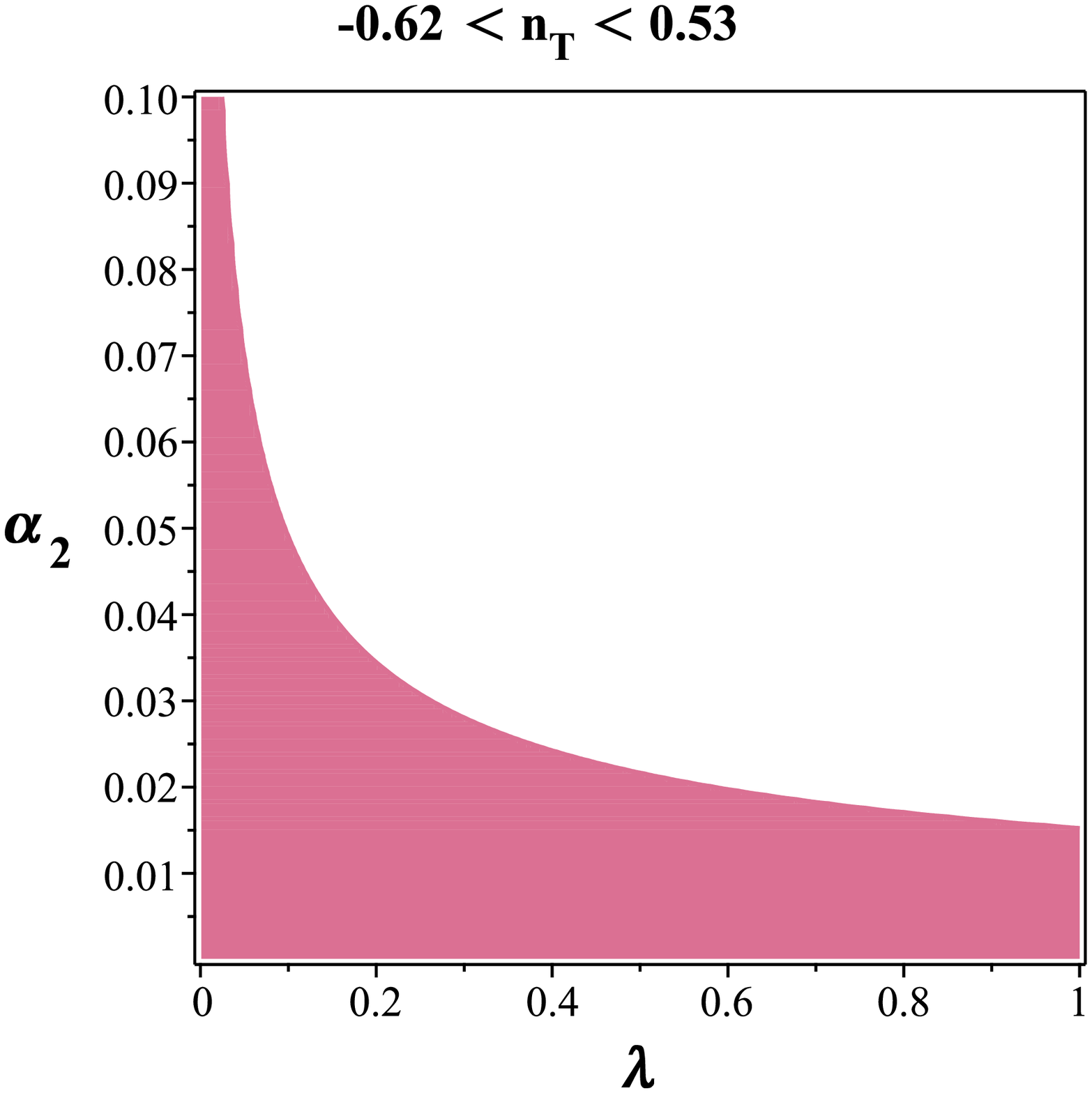}\vspace{8cm}
\end{center}
\caption{\label{fig1}\small {Ranges of the parameters $\alpha_{2}$
and $\lambda$ leading to the observationally viable values of the
scalar spectral index (upper-left panel), tensor-to-scalar ratio
(upper-right panel) and the tensor spectral index (lower panel) in
the tachyon model with the superpotential.}}
\end{figure}

\section{Observational Viability}
To study the observational viability of this model and find some
constraints on the model's parameters, we use the superpotential
given in the equation (\ref{eq22}). By this superpotential, we
obtain slow-roll parameters (\ref{eq18})-(\ref{eq20}) and therefore
the perturbations parameters (scalar spectral index, tensor spectral
index, and tensor-to-scalar ratio) in terms of the scalar fields and
model's parameters. Also, by using equations
(\ref{eq15})-(\ref{eq17}), we find the value of $\phi$ in terms of
the e-folds number and model's parameters and substitute it in the
perturbations parameters. Now, we can compare the values of these
parameters with observational data and obtain some constraints on
the parameter $\alpha_{2}$ for some values of $\lambda$. As we have
mentioned in the Introduction, from Planck2018 TT, TE,
EE+lowE+lensing+BAO+BK14 data and by considering $\Lambda
CDM+r+\frac{dn_{s}}{d\ln k}$ model, we have $n_{s}=0.9658\pm0.0038$
at $68\%$ CL~\cite{pl18a,pl18b}. By using this constraint, we have
obtained the range of the parameters $\alpha_{2}$ and $\lambda$
leading to the observationally viable values of the scalar spectral
index. The result is shown in the upper-left panel of figure 1. The
upper-right panel of figure 1 shows the range of the parameters
$\alpha_{2}$ and $\lambda$ which leads to $r<0.072$, obtained from
Planck2018 TT, TE, EE+lowE+lensing+BAO+BK14 data at $68\%$ CL. Also,
the lower panel of figure 1 shows the range of the parameters
$\alpha_{2}$ and $\lambda$ leading to $-0.62<n_{T}<0.53$, obtained
from Planck2018 TT, TE, EE +lowE+lensing+BK14+BAO+LIGO and Virgo2016
data at $68\%$ CL~\cite{pl18a,pl18b}. To plot these figures, we have
adopted $N=60$, $0<\alpha_{2}\leq 0.1$ and $0<\lambda\leq 1$.

To obtain some constraints on the model's parameters space, we have
plotted the evolution of the tensor-to-scalar ratio in the
background of the Planck2018 TT, TE, EE+lowE+lensing+BAO+BK14 data.
The result is shown in figure 2. As the figure shows, the tachyon
inflation with superpotential in some ranges of the model's
parameter is consistent with observational data. We also have
plotted the evolution of the tensor-to-scalar ratio versus the
tensor spectral index in the background of the Planck2018 TT, TE, EE
+lowE+lensing+BK14+BAO+LIGO and Virgo2016 data set in figure 3. Note
that, as equation (\ref{eq35}) and figure 3 show, the consistency
relation in this setup is satisfied. By Using the numerical
analysis, we have obtained some constraints on the parameter
$\alpha_{2}$, for some sample values of $\lambda$ as $0.1$, $0.3$,
$0.6$ and $0.9$, which have been summarized in table 1. Also, by
using the constraints that Planck2018 TT, TE,
EE+lowE+lensing+BAO+BK14 data at $68\%$ CL and $95\%$ CL set on the
scalar spectral index and the tensor-to-scalar ratio, it is possible
to obtain the ranges of the sound speed and $\lambda$ which fulfill
the mentioned data. In this respect, figure 4 shows the
observationally viable ranges of $\lambda$ and $c_{s}$ at $68\%$ CL
(yellow region) and $95\%$ CL (magenta region).

\begin{figure}
\begin{center}\includegraphics{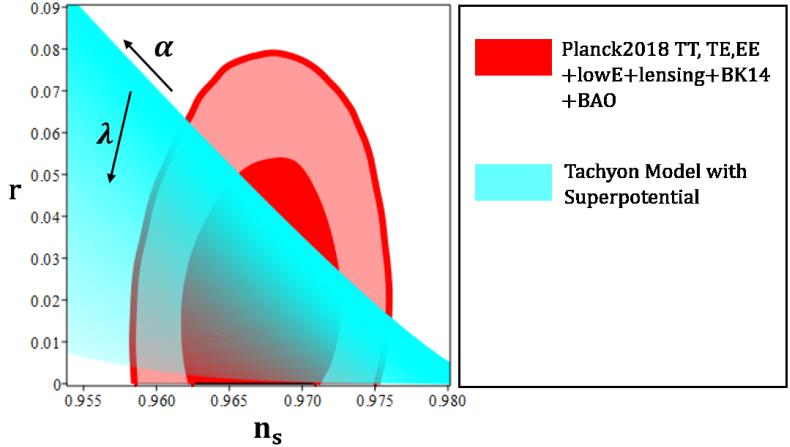}\vspace{6cm}
\end{center}
\caption{\label{fig2}\small {Tensor-to-scalar ratio versus the
scalar spectral index in the tachyon model with superpotential. The
arrows show the direction in which the parameters increase.}}
\end{figure}

\begin{figure}
\begin{center}\includegraphics{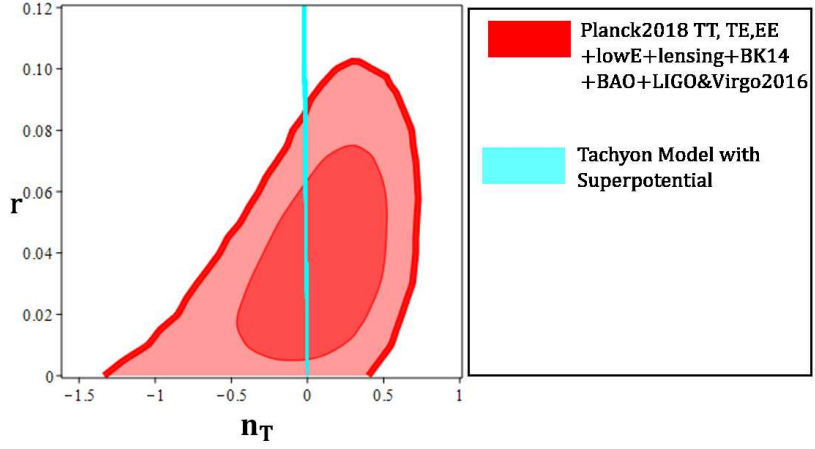}\vspace{6cm}
\end{center}
\caption{\label{fig3}\small {Tensor-to-scalar ratio versus the
tensor spectral index in the tachyon model with superpotential.}}
\end{figure}

\begin{table*}
\tiny \tiny \caption{\small{\label{tab:1} Ranges of the parameter
$\alpha_{2}$ in which the tensor-to-scalar ratio, the scalar
spectral index and the tensor spectral index of the tachyon
inflation with superpotential are consistent with different data
sets.}}
\begin{center}
\begin{tabular}{cccccc}
\\ \hline \hline \\ & Planck2018 TT,TE,EE+lowE & Planck2018 TT,TE,EE+lowE&Planck2018 TT,TE,EE+lowE&Planck2018 TT,TE,EE+lowE
\\
& +lensing+BK14+BAO &
+lensing+BK14+BAO&lensing+BK14+BAO&lensing+BK14+BAO
\\
&  & &+LIGO$\&$Virgo2016 &LIGO$\&$Virgo2016
\\
\hline \\$\lambda$& $68\%$ CL & $95\%$ CL &$68\%$ CL & $95\%$ CL
\\
\hline\hline \\  $0.1$& $1.41\times 10^{-2}<\alpha_{2}<1.83\times
10^{-2}$ &$1.11\times 10^{-2}<\alpha_{2}<1.98\times 10^{-2}$
&$0.65\times 10^{-2}<\alpha_{2}<2.0.2\times 10^{-2} $ & $0<\alpha_{2}<2.25\times 10^{-2}$\\ \\
\hline
\\$0.4$&$0.68\times 10^{-2}<\alpha_{2}<0.91\times
10^{-2}$ &$0.55\times 10^{-2}<\alpha_{2}<0.98\times 10^{-2}$
&$0.35\times 10^{-2}<\alpha_{2}<1.03\times 10^{-2}$
&$0<\alpha_{2}<1.15\times 10^{-2}$
\\ \\ \hline\\
$0.7$&$0.48\times 10^{-2}<\alpha_{2}<0.68\times 10^{-2}
$&$0.39\times 10^{-2}<\alpha_{2}<0.73\times 10^{-2}$&$0.28\times
10^{-2}<\alpha_{2}<0.83\times
10^{-2} $ &$0<\alpha_{2}<0.92\times 10^{-2}$\\ \\
\hline\\
$0.9$&$0.17\times 10^{-2}<\alpha_{2}<0.55\times 10^{-2}
$&$0.12\times 10^{-2}<\alpha_{2}<0.40\times 10^{-2}$&
$0.43\times 10^{-2}<\alpha_{2}<0.84\times 10^{-2} $ &$0<\alpha_{2}<0.90\times 10^{-2}$\\ \\
\hline \hline
\end{tabular}
\end{center}
\end{table*}

\begin{figure}
\begin{center}\includegraphics{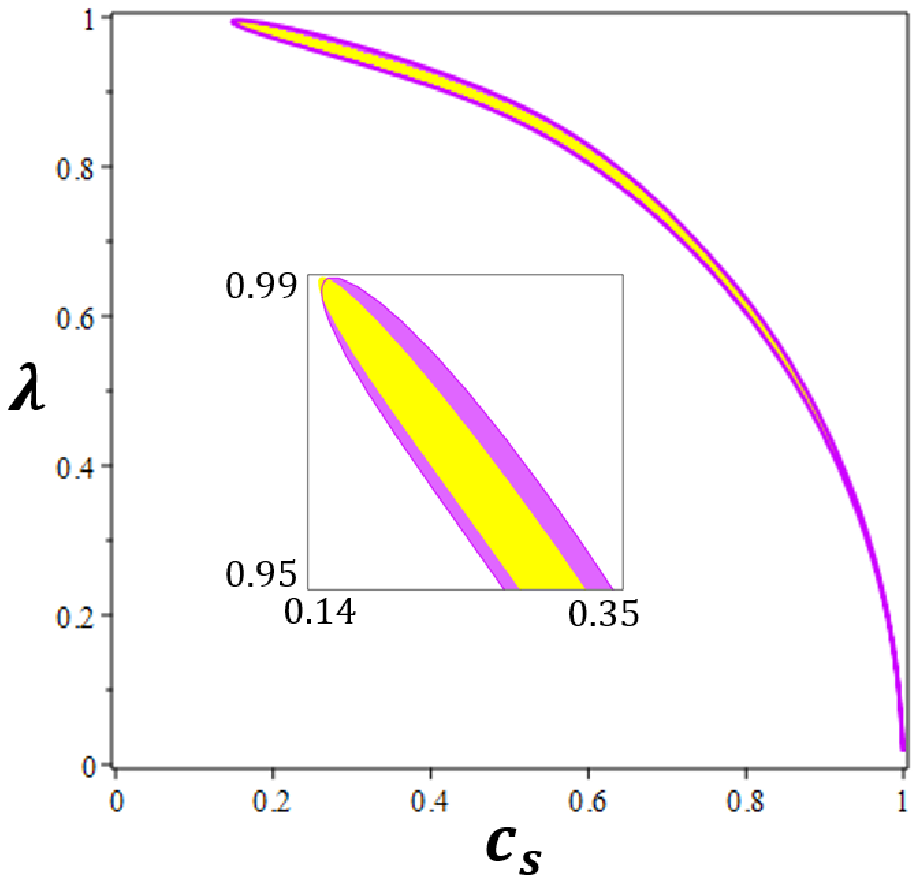}\vspace{6cm}
\end{center}
\caption{\label{fig4}\small {Ranges of the parameters $\lambda$ and
$c_{s}$ leading to the observationally viable values of the scalar
spectral index and the tensor-to-scalar ratio, obtained from
Planck2018 TT, TE, EE+lowE+lensing+BAO+BK14 data at $68\%$ CL
(yellow region) and $95\%$ CL (magenta region).}}
\end{figure}

To study the non-gaussian feature of the primordial perturbations
numerically, we consider equation (\ref{eq44}) with the sound speed
given in equation (\ref{eq21}) and the superpotential (\ref{eq22}).
In fact, we seek for the prediction of the model about the amplitude
of the non-gaussianity in the tachyon model with superpotential. In
this regard, in figure 6 we have plotted the phase space of the
parameters $\alpha_{2}$ and $\lambda$ and also the corresponding
predicted values of the amplitude of the equilateral configuration
of the non-gaussianity.  As the figure shows, for most ranges of the
parameter space, the tachyon model with superpotential predicts
small non-gaussianity. In fact, if we consider $\lambda \leq 0.9$,
for all values of $\alpha_{2}$, the amplitude of the equilateral
non-gaussianity in the tachyon model is consistent with Planck2018
TTT, EEE, TTE and EET data at $68\%$ CL. However, if we consider
$\lambda > 0.9$, there would be some constraints on the parameter
$\alpha_{2}$ in confrontation with Planck2018 TTT, EEE, TTE and EET
data at $68\%$ CL. We have obtained some of these constraints which
have been summarized in table 2. Also, from the constraints obtained
by comparing the $r-n_{s}$ and $r-n_{T}$ behavior in the background
of Planck2018 TT, TE, EE+lowE+lensing+BAO+BK14 data, it is possible
to find the observationally viable values of the sound speed and the
amplitude of the equilateral configuration of the non-gaussianity.
The results of this numerical analysis have been summarized in table
3.

\begin{figure}
\begin{center}\includegraphics{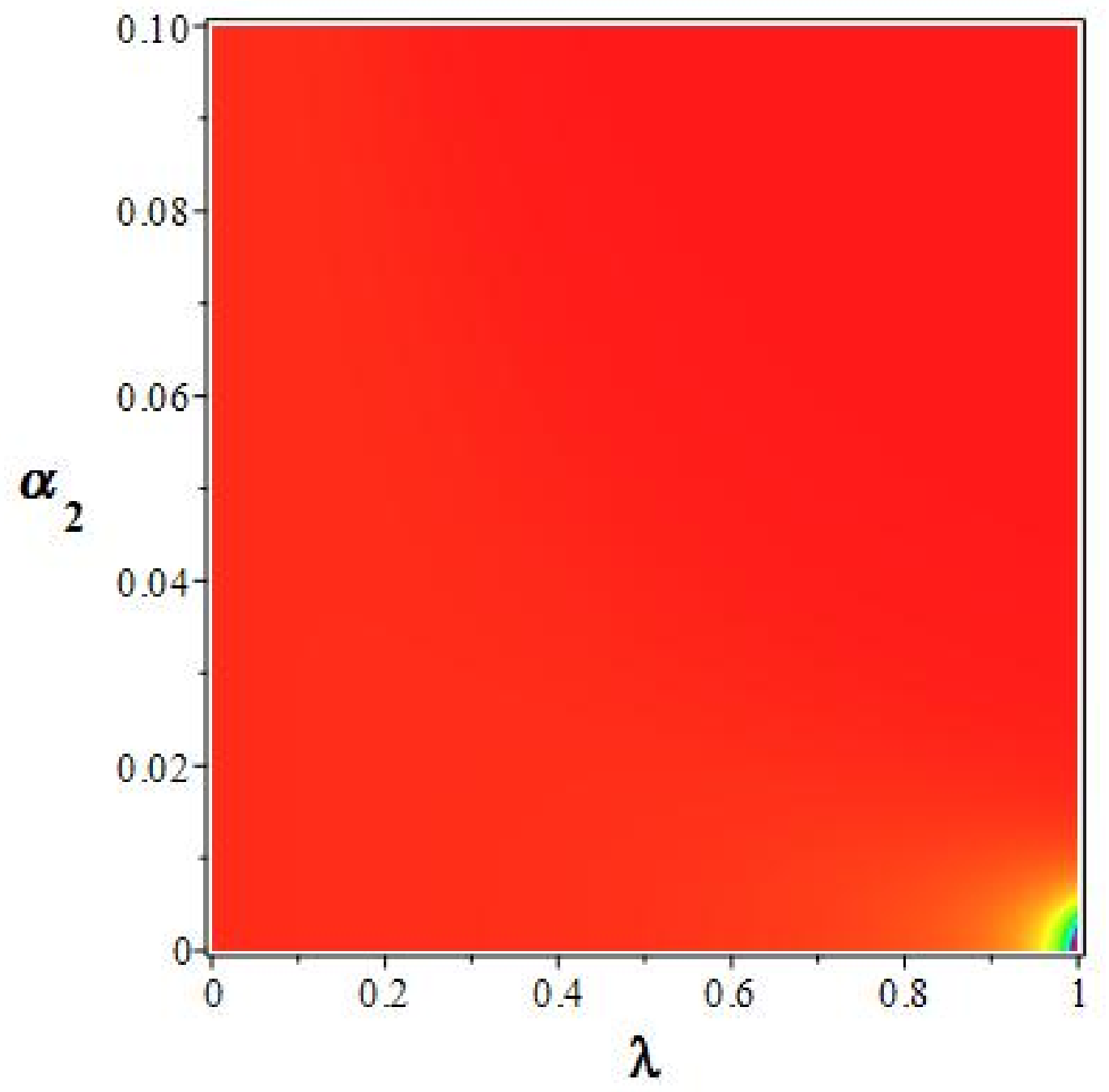}\includegraphics{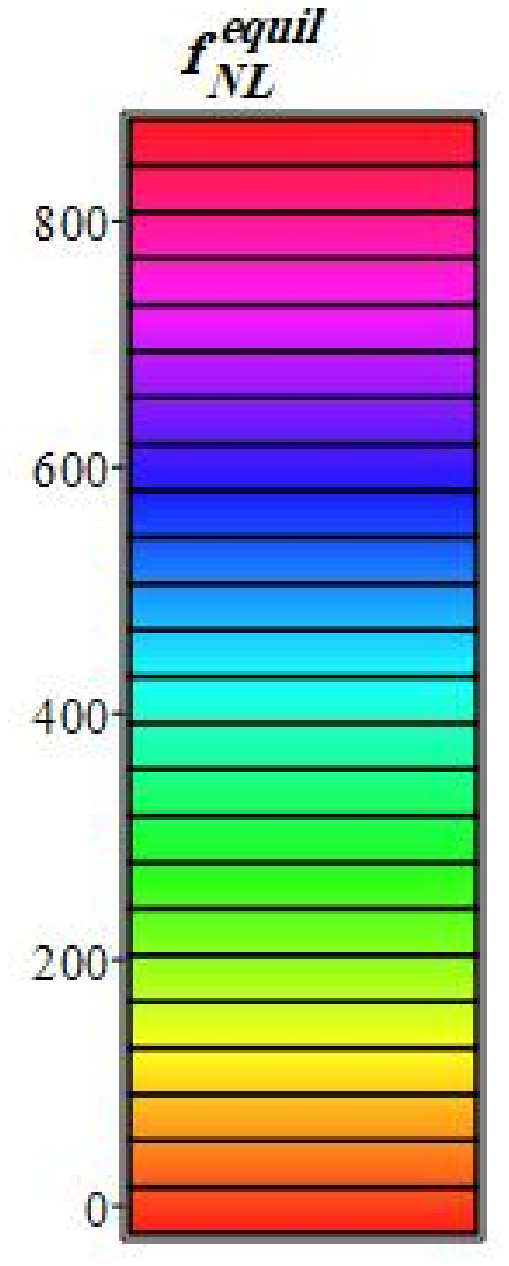}\vspace{7.5cm}
\end{center}
\caption{\label{fig5}\small {Space of the parameters $\alpha_{2}$
and $\lambda$ (left panel) and the corresponding predicted values of
the amplitude of the equilateral non-gaussianity (right panel) in
the tachyon model with superpotential.}}
\end{figure}

\begin{table*}
\caption{\small{\label{tab:2} Ranges of the parameters $\alpha_{2}$
and $\lambda$ leading to the observationally viable values of the
equilateral configuration of the non-gaussianity in the tachyon
model with superpotential, in confrontation with the Planck2018 TTT,
EEE, TTE and EET data at $68\%$ CL.}}
\begin{center}
\begin{tabular}{cccccc}
\\ \hline \hline $0<\lambda\leq 0.9$ & $\lambda=0.93$  & $\lambda=0.96$  &$\lambda=0.99$
\\
\hline\hline \\ $0<\alpha_{2}<0.1$ & $0.004<\alpha_{2}<0.1 $&$
0.006<\alpha_{2}<0.1$&$0.008<\alpha_{2}<0.1$ \\
\hline \hline
\end{tabular}
\end{center}
\end{table*}

\begin{table*}
\small \caption{\small{\label{tab:3} The observationally viable
ranges of the sound speed and the amplitude of the equilateral
configuration of the non-gaussianity in the tachyon inflation with
superpotential, in confrontation with Planck2018 TT, TE,
EE+lowE+lensing+BAO+BK14 data at $68\%$ CL.}}
\begin{center}
\begin{tabular}{cccccccc}
\\ \hline \hline  \\$\lambda$&& $c_{s}$  && $f_{NL}^{equil}$
\\
\hline\hline
\\$0.1$&&$0.994<c_{s}< 0.995$&&$0.039<f_{NL}^{equil}<0.040$
\\ \\ \hline\\
$0.4$&& $0.920<c_{s}< 0.923 $ &&$0.727 <f_{NL}^{equil}<0.759$\\ \\
\hline\\
$0.7$&& $0.727 <c_{s}< 0.740 $ &&$3.47 <f_{NL}^{equil}< 3.74$\\ \\
\hline\\
$0.9$&& $0.464  <c_{s}< 0.493 $ &&$13.0 <f_{NL}^{equil}< 15.2$\\ \\
\hline \hline
\end{tabular}
\end{center}
\end{table*}

\section{Reheating Phase after Inflation}

After ending the inflationary expansion, the universe should be
reheated for subsequent evolution. This process is named reheating
phase of the universe which studying it gives us some more
information about the model. In this regard,
following~\cite{Dai14,Un15,Co15,Cai15,Ue16}, in this section we
obtain some expressions for the e-folds number during the reheating,
$N_{rh}$, (the subscript ``$rh$" stands for the reheating) and
temperature during this era, $T_{rh}$. Considering that these
reheating parameters are expressed in terms of the scalar spectral
index, it is possible to explore their observational viability.

We start with the following equation
\begin{equation}
\label{eq45} N_{hc}=\ln \left(\frac{a_{e}}{a_{hc}}\right)\,,
\end{equation}
defining the e-folds number between the time where physical scales
cross the horizon and the time where the inflation epoch ends. By
the subscript ``$e$'' we mean the end of inflation and by the
subscript ``$hc$'' we show the horizon crossing. Note that, the
relation $\rho\sim a^{-3(1+\omega_{eff})}$ for the energy density is
satisfied during the reheating process. In this relation, the
effective equation of state, corresponding to the dominant energy
density of the universe, is shown by the parameter $\omega_{eff}$.
By using this relation between the energy density and the equation
of state parameter, we can find the e-folds number during reheating
as follows
\begin{eqnarray}\label{eq46}
N_{rh}=\ln\left(\frac{a_{rh}}{a_{e}}\right)=-\frac{1}{3(1+\omega_{eff})}\ln\left(\frac{\rho_{rh}}{\rho_{e}}\right)\,.
\end{eqnarray}
On the other hand, at the horizon crossing of the physical scales we
have
\begin{eqnarray}\label{eq47}
0=\ln\left(\frac{k_{hc}}{a_{hc}H_{hc}}\right)=
\ln\left(\frac{a_{e}}{a_{hc}}\frac{a_{rh}}{a_{e}}\frac{a_{0}}{a_{rh}}\frac{k_{hc}}{a_{0}H_{hc}}\right)\,,
\end{eqnarray}
with subscript ``$0$'' being the current value of the corresponding
parameter. Now, equations (\ref{eq45})-(\ref{eq47}) give
\begin{eqnarray}\label{eq48}
N_{hc}+N_{rh}+\ln\left(\frac{k_{hc}}{a_{0}H_{hc}}\right)+\ln\left(\frac{a_{0}}{a_{rh}}\right)=0\,.
\end{eqnarray}
Now, we express $\frac{a_{0}}{a}$ in terms of the temperature and
density. To this end, we use the following relation between the
energy density and temperature during the reheating~\cite{Co15,Ue16}
\begin{equation}\label{eq49}
\rho_{rh}=\frac{\pi^{2}g_{rh}}{30}T_{rh}^{4}\,.
\end{equation}
In the above equation, the parameter $g_{rh}$ is the effective
number of the relativistic species at the reheating era. Also, the
conservation of the entropy gives the following equation
\cite{Co15,Ue16}
\begin{equation}\label{eq50}
\frac{a_{0}}{a_{rh}}=\left(\frac{43}{11g_{rh}}\right)^{-\frac{1}{3}}\frac{T_{rh}}{T_{0}}\,.
\end{equation}
From equations (\ref{eq49}) and (\ref{eq50}) we get
\begin{eqnarray}\label{eq51}
\frac{a_{0}}{a_{rh}}=\left(\frac{43}{11g_{rh}}\right)^{-\frac{1}{3}}T_{0}^{-1}\left(\frac{\pi^{2}g_{rh}}{30\rho_{rh}}\right)^{-\frac{1}{4}}\,.
\end{eqnarray}

In the tachyon model with superpotential, the energy density in
terms of the slow-roll parameter $\epsilon$ is given by
\begin{eqnarray}\label{eq52}
\rho=\frac{\frac{\kappa^{2}}{6}W^{2}-\frac{1}{8}W'^{2}}{\sqrt{1-\frac{2}{3}\epsilon}}\,.
\end{eqnarray}
At the end of inflation, where we have $\epsilon=1$, the energy
density takes the following form
\begin{equation}\label{eq53}
\rho_{e}=\sqrt{3}\,\bigg(\frac{\kappa^{2}}{6}W_{e}^{2}-\frac{1}{8}W_{e}'^{2}\bigg)\,.
\end{equation}
The energy density during the reheating era is obtained from
equations (\ref{eq46}) and (\ref{eq53}) as follows
\begin{eqnarray}\label{eq54}
\rho_{rh}=\sqrt{3}\,\bigg(\frac{\kappa^{2}}{6}W_{e}^{2}-\frac{1}{8}W_{e}'^{2}\bigg)\,\exp\Big[-3N_{rh}(1+\omega_{eff})\Big].
\end{eqnarray}
Now, equations (\ref{eq51}) and (\ref{eq53}) give
\begin{eqnarray}\label{eq55}
\ln\left(\frac{a_{0}}{a_{rh}}\right)=-\frac{1}{3}\ln\left(\frac{43}{11g_{rh}}\right)
-\frac{1}{4}\ln\left(\frac{\pi^{2}g_{rh}}{30\rho_{rh}}\right)-\ln
T_{0}
+\frac{1}{4}\ln\Bigg(\sqrt{3}\,\bigg(\frac{\kappa^{2}}{6}W_{e}^{2}-\frac{1}{8}W_{e}'^{2}\bigg)\Bigg)\nonumber\\
-\frac{3}{4}N_{rh}(1+\omega_{eff})\,.
\end{eqnarray}
By using equations (\ref{eq29}), (\ref{eq48}) and (\ref{eq55}), we
find the e-folds number during reheating as follows
\begin{eqnarray}\label{eq56}
N_{rh}=\frac{4}{1-3\omega_{eff}}\Bigg[-N_{hc}-\ln\Big(\frac{k_{hc}}{a_{0}T_{0}}\Big)-\frac{1}{4}\ln\Big(\frac{40}{\pi^{2}g_{rh}}\Big)
-\frac{1}{3}\ln\Big(\frac{11g_{rh}}{43}\Big)\nonumber\\+\frac{1}{2}\ln\Big(8\pi^{2}{\cal{A}}_{s}{\cal{Q}}_{s}
c_{s}^{3}\Big) -\frac{1}{4}\ln\Bigg(\sqrt{3}\,
\bigg(\frac{\kappa^{2}}{6}W_{e}^{2}-\frac{1}{8}W_{e}'^{2}\bigg)\Bigg)\Bigg].
\end{eqnarray}
We can also find the temperature during reheating from equations
(\ref{eq46}), (\ref{eq50}) and (\ref{eq53}) as follows
\begin{equation}\label{eq57}
T_{rh}=\bigg(\frac{30}{\pi^{2}g_{rh}}\bigg)^{\frac{1}{4}}\,
\bigg[\sqrt{3}\,
\bigg(\frac{\kappa^{2}}{6}W_{e}^{2}-\frac{1}{8}W_{e}'^{2}\bigg)\bigg]^{\frac{1}{4}}\,\exp\bigg[-\frac{3}{4}N_{rh}(1+\omega_{eff})\bigg]\,.
\end{equation}

Now, we study the reheating process numerically. To this end, by
using equation (\ref{eq2}), we obtain equations (\ref{eq56}) and
(\ref{eq57}) in terms of the tachyon field $\phi$. After that, we
use equations (\ref{eq18}), (\ref{eq19}), (\ref{eq20}) and
(\ref{eq31}) to find the value of the tachyon field in terms of the
scalar spectral index. In this way, we get the e-folds number and
the temperature during the reheating phase in terms of $n_{s}$. By
using the observational constraint on the scalar spectral index,
obtained from Planck2018 TT, TE, EE+lowE+lensing+BK14+BAO joint
data, we can find some constraints on $n_{rh}$ and $T_{rh}$. In this
regard, in figure 6 we have plotted the e-fold number during
reheating versus the effective equation of state. This figure has
been plotted for some sample values of the parameters $\lambda$ and
$\alpha_{2}$ which have been adopted from the constraints presented
in table 1. Also, to plot this figure we have used the constraint on
the scalar spectral index as $n_{s}=0.9658\pm0.0038$. The range used
for the effective equation of state is $-1\leq\omega_{eff}\leq1$. In
fact, at the end of the inflationary expansion, we have
$\omega_{eff}=-\frac{1}{3}$ and at the beginning of the radiation
dominated era $\omega_{eff}=\frac{1}{3}$. Therefore, the adopted
range for the effective equation of state seems suitable for
consideration. As the figure shows, the value of e-folds number
during reheating increases by changing the value of the effective
equation of state from $-1$ to $\frac{1}{3}$. Also, our numerical
analysis shows that for $\lambda<0.2$ and $\alpha_{2}<1.53\times
10^{-2}$, it is possible to have instantaneous reheating. However,
for $\lambda\geq0.2$ and $\alpha_{2}\geq 1.53\times 10^{-2}$, a few
e-folds number are needed for the reheating process to be completed.
We have also plotted the behavior of $N_{rh}$ versus $n_{s}$, for
some sample values of $\lambda$ and $\alpha_{2}$, in figure 7. In
this figure, we have also adopted three values of the effective
equation of state as $\omega_{eff}=-1,-\frac{1}{3}$ and $0$. The
behavior of the temperature during the reheating versus the scalar
spectral index is shown in figure 8. This figure has been plotted
for the same values of the parameters as figure 7. By performing
these numerical analyses, we have obtained some constraints on the
important reheating parameters $N_{rh}$ and $T_{rh}$, summarized in
tables 4 and 5.

\begin{figure}
\begin{center}\includegraphics{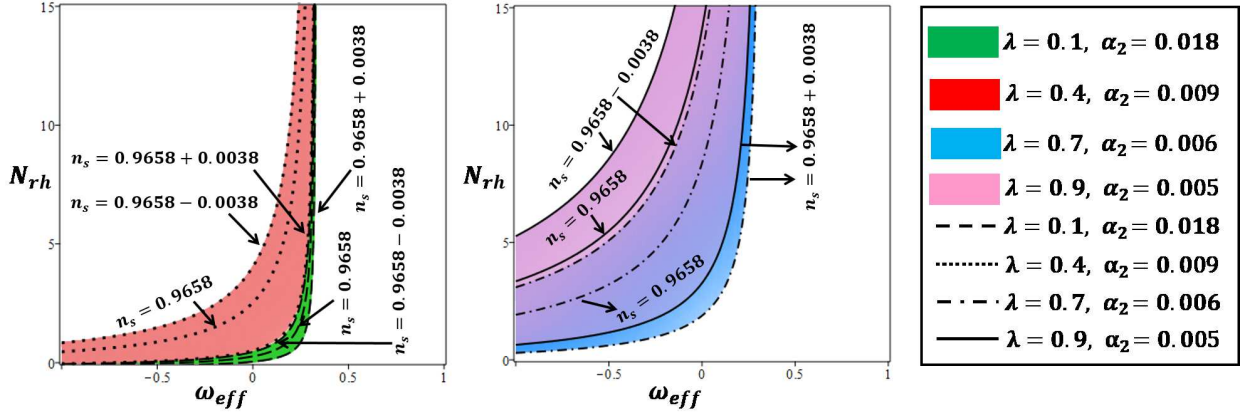}\vspace{6cm}
\end{center}
\caption{\label{fig6}\small {Ranges of the e-folds number and the
effective equation of state parameter during the reheating phase for
the tachyon model with superpotential, leading to the
observationally viable values of the scalar spectral index. }}
\end{figure}

\begin{figure}
\begin{center}\includegraphics{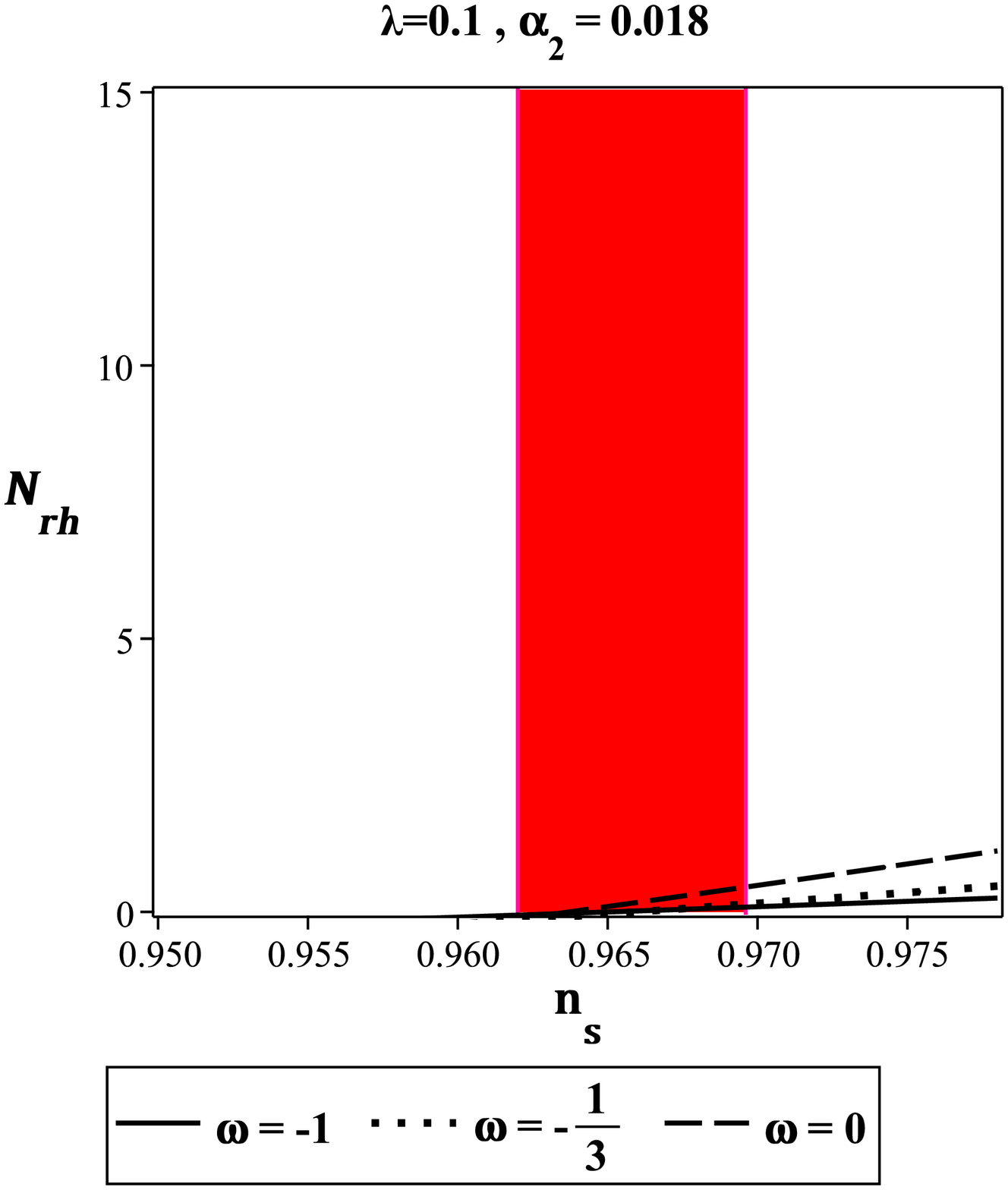}\includegraphics{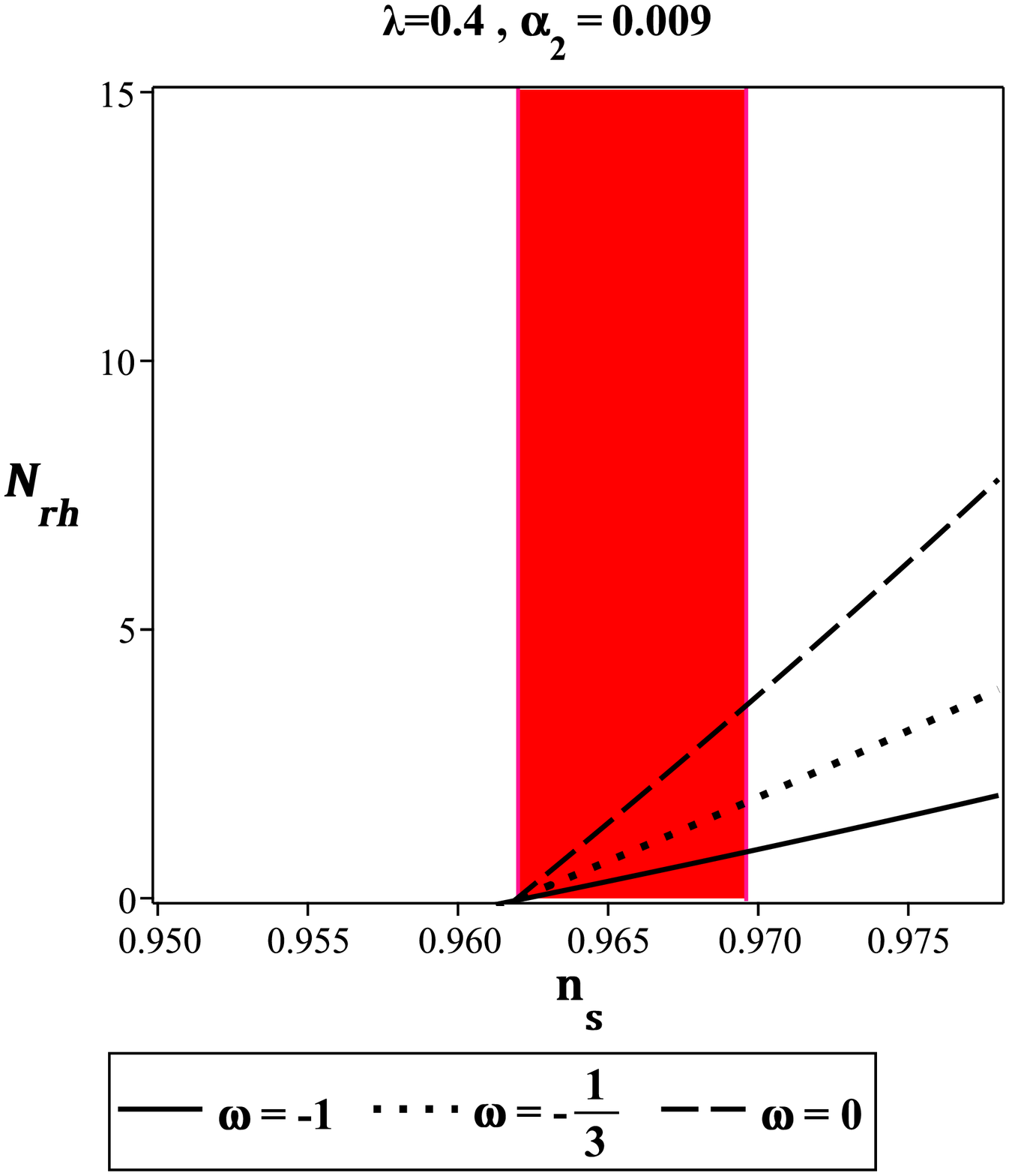}\vspace{10.3cm}
\includegraphics{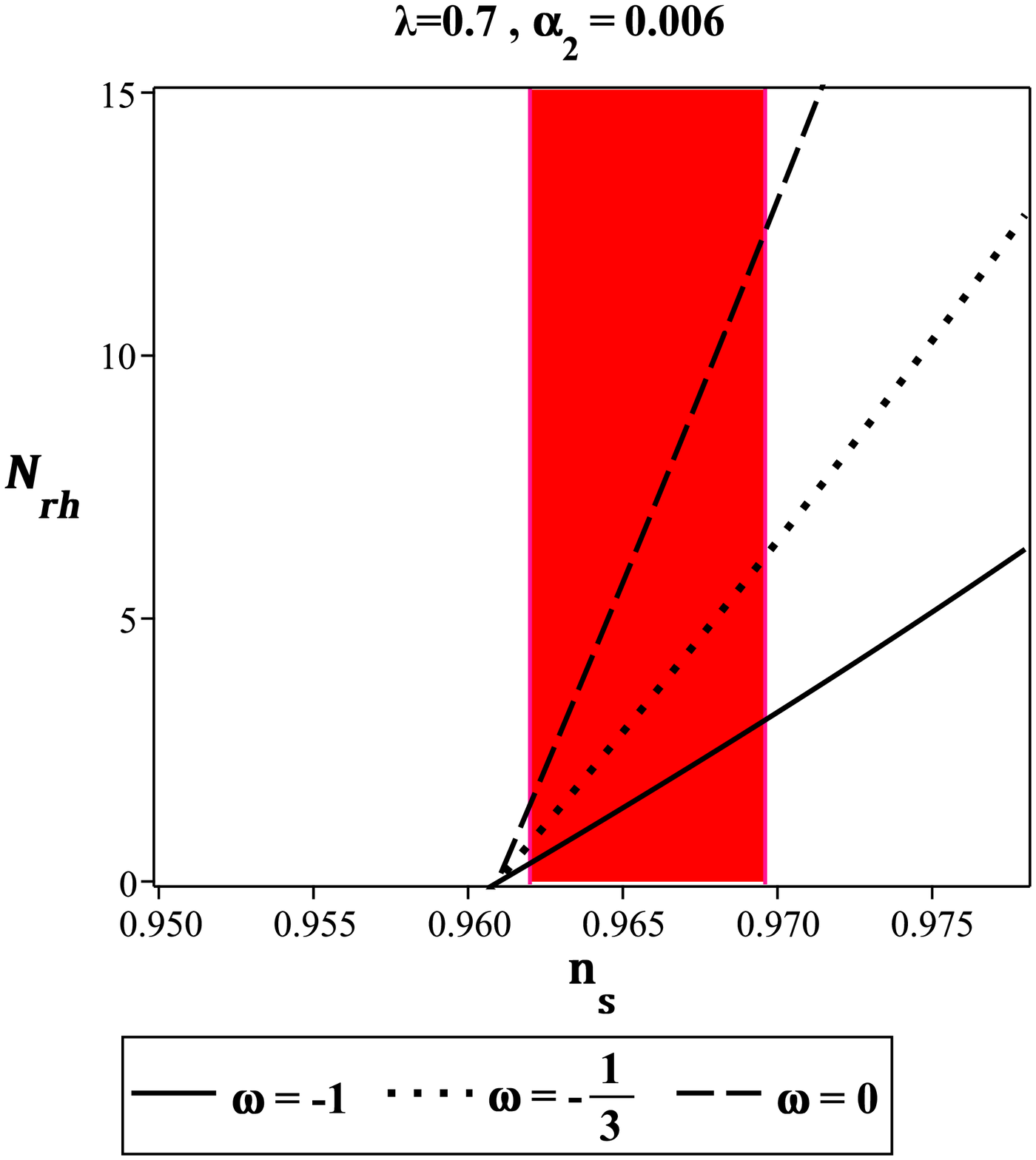}\includegraphics{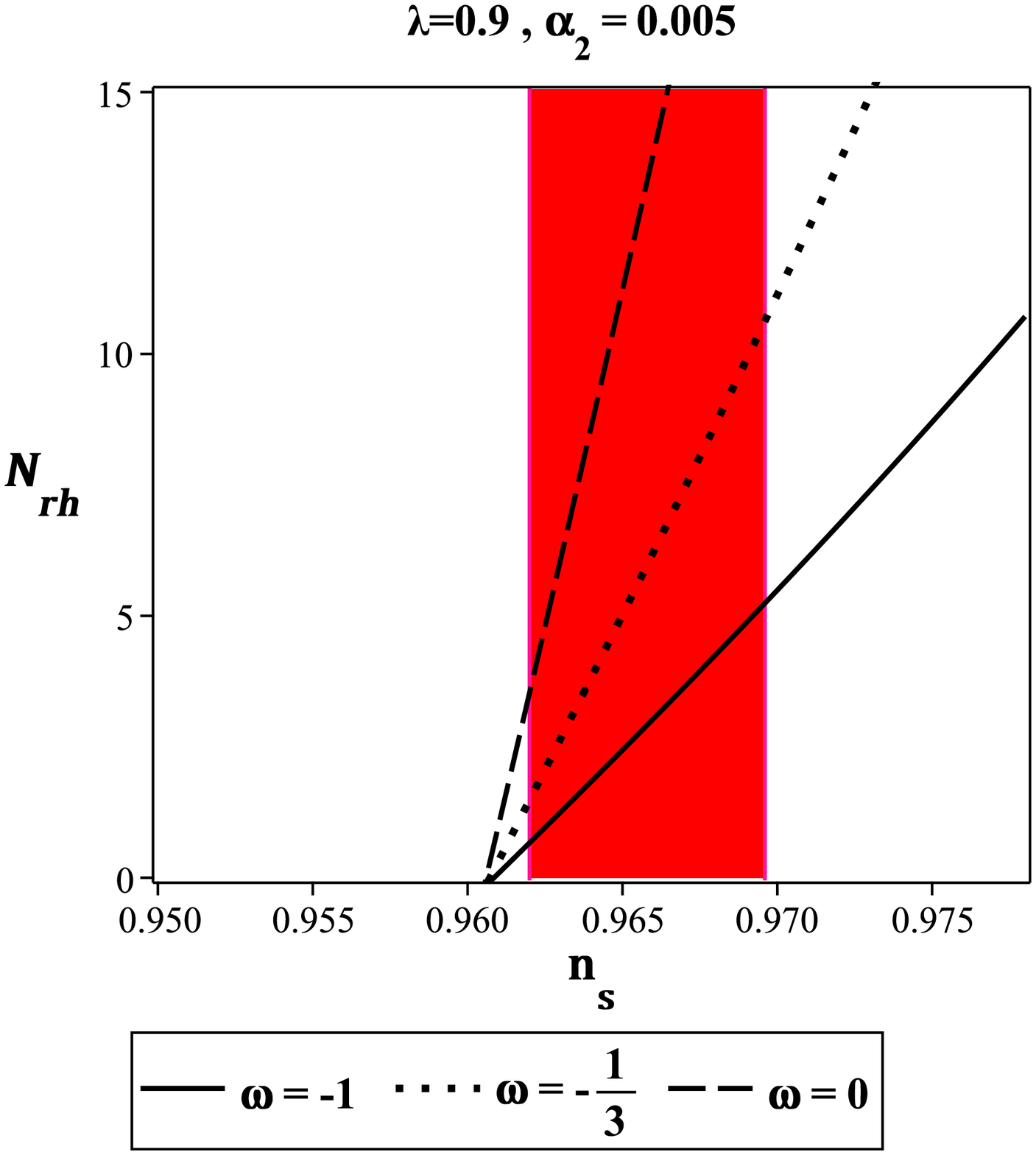}\vspace{7cm}
\end{center}
\caption{\label{fig7}\small {The behavior of the e-folds number
during the reheating phase versus the scalar spectral index, for the
tachyon model with superpotential. The red region shows the values
of the scalar spectral index released by Planck2018 TT, TE,
EE+lowE+lensing+BK14+BAO joint data at $68\%$ CL.}}
\end{figure}

\begin{figure}
\begin{center}\includegraphics{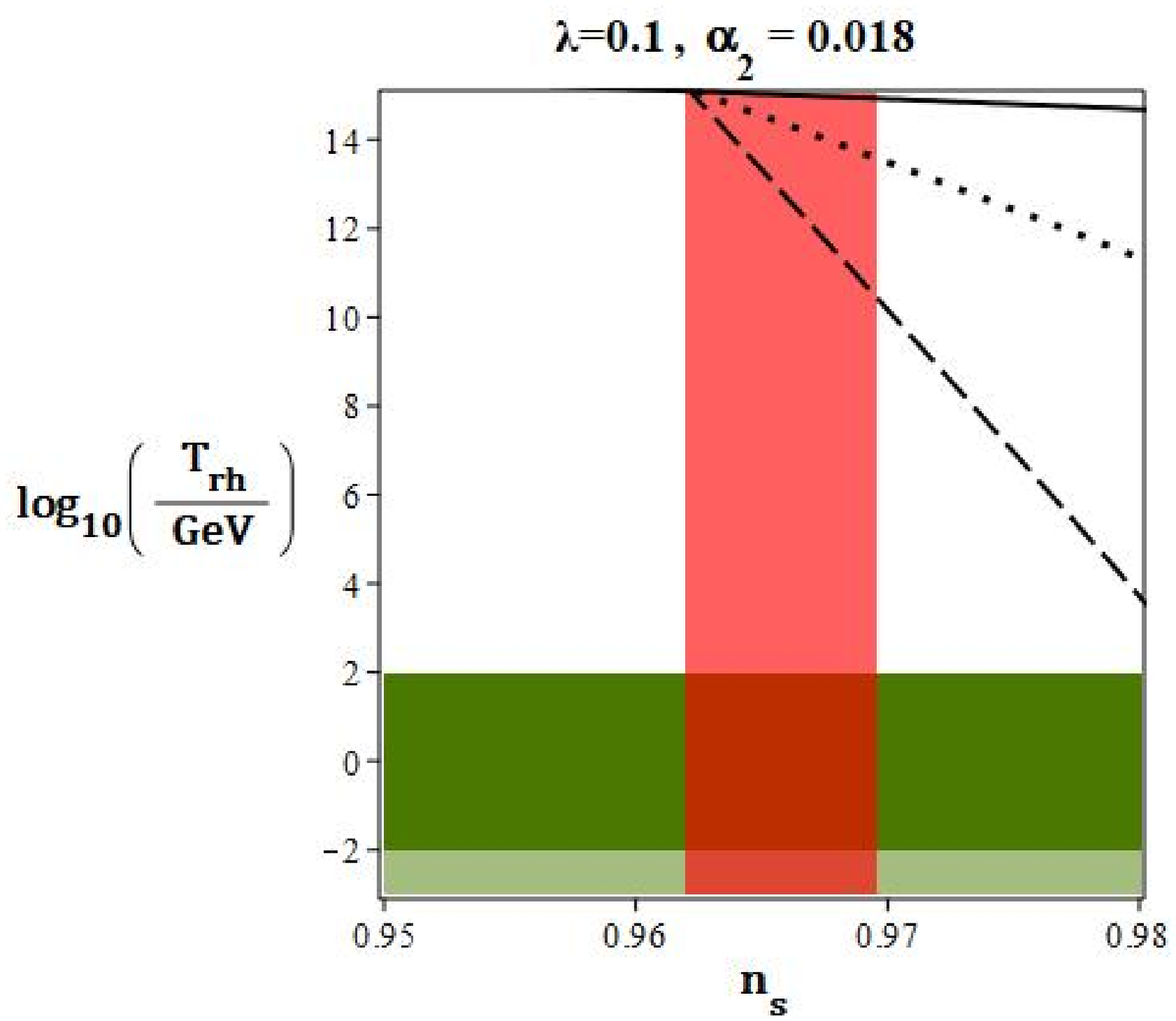}\includegraphics{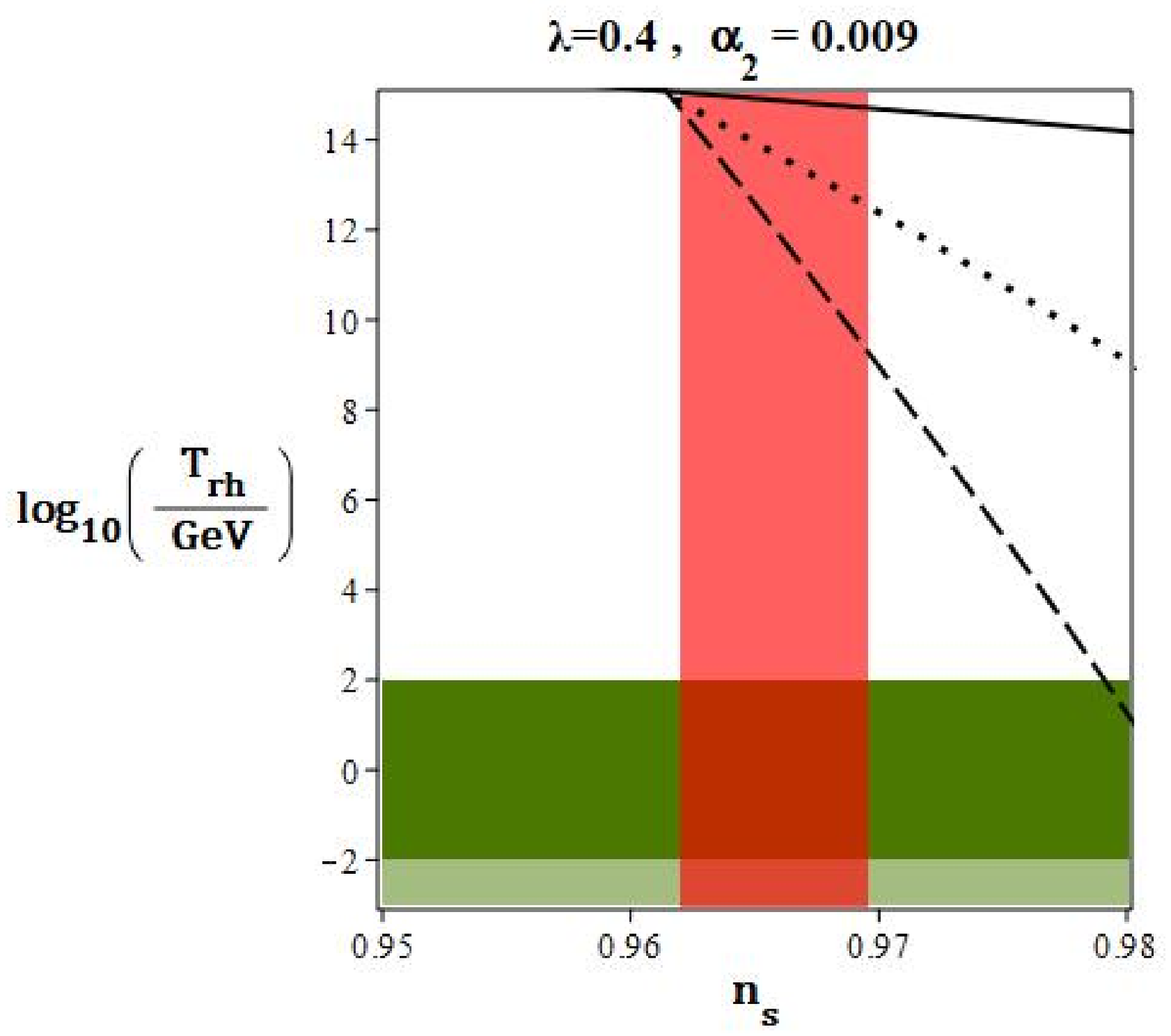}\vspace{7cm}
\includegraphics{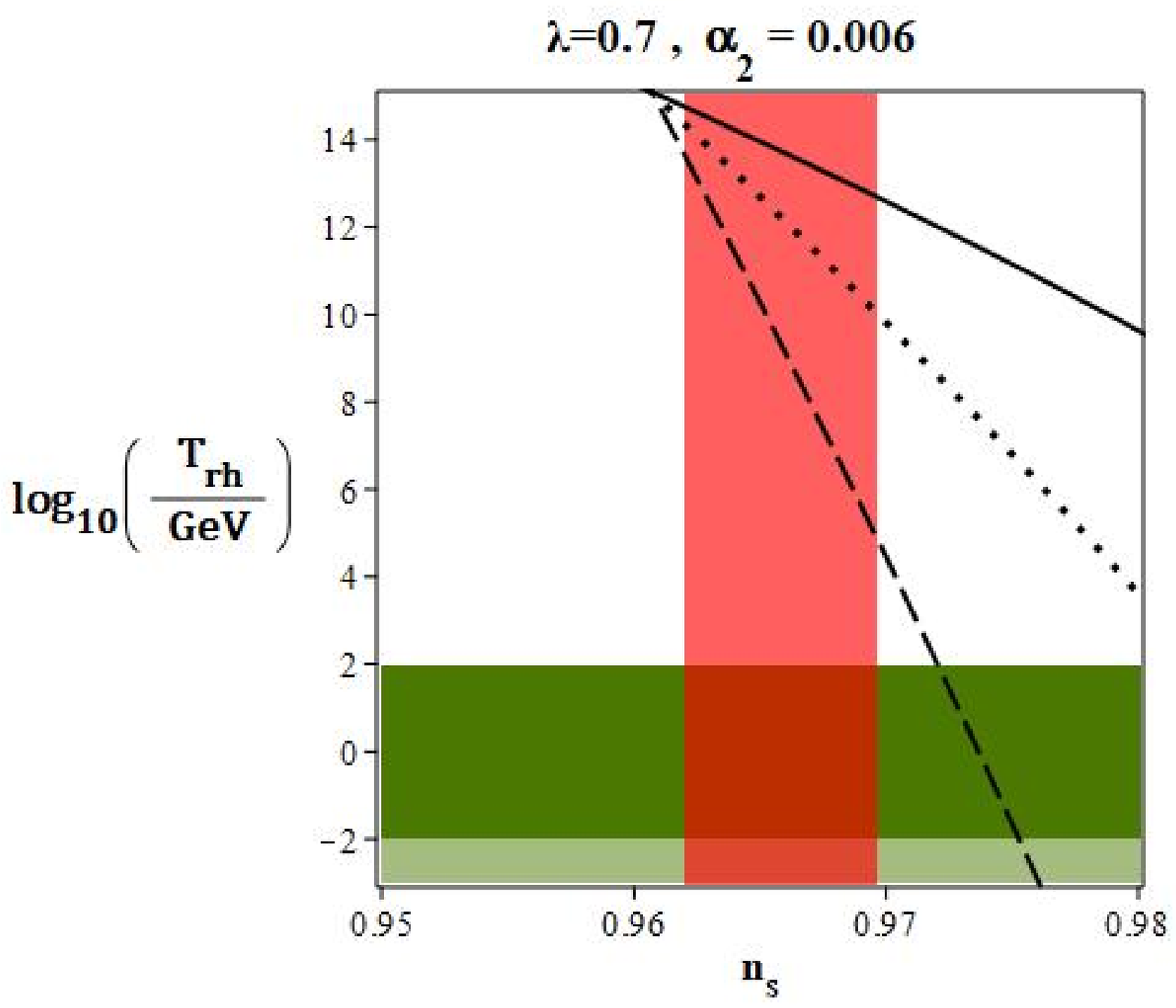}\includegraphics{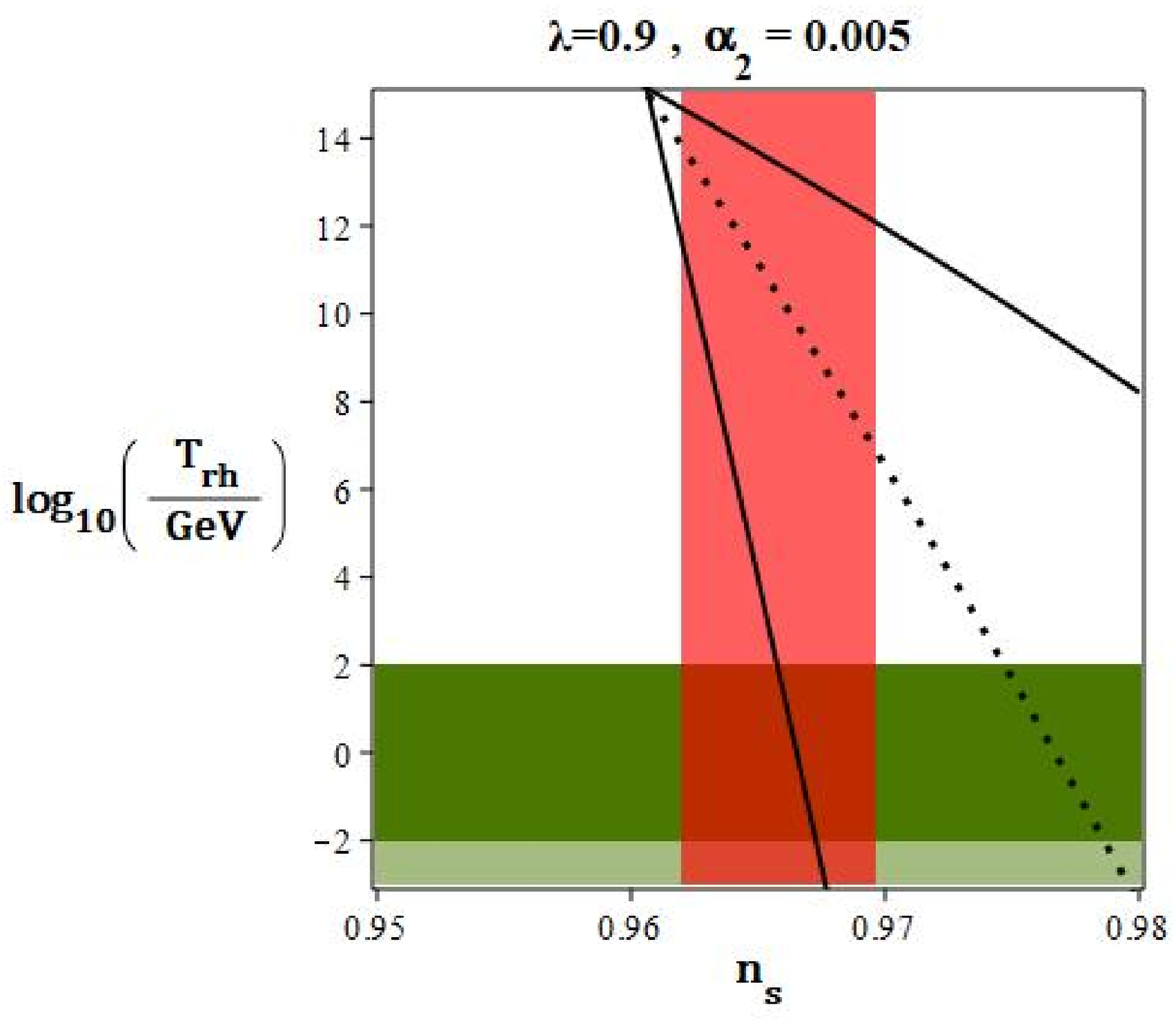}\vspace{6cm}
\end{center}
\caption{\label{fig8}\small {The behavior of the temperature during
the reheating phase versus the scalar spectral index, for the
tachyon model with superpotential. The dark green region shows the
temperatures below the electroweak scale, $T<100$ GeV, and the light
green region corresponds to the temperatures below the big bang
nucleosynthesis scale, $T<10$ MeV.}}
\end{figure}

\begin{table*}
\tiny \tiny \caption{\small{\label{tab:4} Constraints on the e-folds
number during reheating in the tachyon model with superpotentail,
obtained from Planck2018 TT, TE, EE+lowE+lensing+BK14+BAO joint data
at $68\%$ CL.}}
\begin{center}
\begin{tabular}{cccccc}
\\ \hline \hline \\ $\lambda$&$\alpha_{2}$& $\omega=-1$& $\omega=-\frac{1}{3}$
&$\omega=0$
\\
\hline
\\$0.1$&$1.41\times10^{-2}<\alpha_{2}<1.83\times10^{-2}$&$0\leq N_{rh}\leq0.134$&$0\leq N_{rh}\leq0.181$&$0.012\leq N_{rh}\leq0.513$\\ \\
\hline
\\$0.4$&$0.68\times 10^{-2}<\alpha_{2}<0.91\times
10^{-2}$&$0.010\leq N_{rh}\leq 0.883$&$0.197 \leq N_{rh}\leq 1.22$&$0.514 \leq N_{rh}\leq 4.31$\\ \\
\hline
\\$0.7$&$0.48\times 10^{-2}<\alpha_{2}<0.68\times 10^{-2}$&$0.201 \leq N_{rh}\leq 3.32$&$0.862\leq N_{rh}\leq 6.67$&
$1.72\leq N_{rh}\leq 15.6$\\ \\
\hline
\\$0.9$&$0.17\times 10^{-2}<\alpha_{2}<0.33\times 10^{-2}
$&$0.541\leq N_{rh}\leq 5.81$&$1.404 \leq N_{rh}\leq 7.66$&
$2.89 \leq N_{rh}\leq 22.2$\\ \\
\hline \hline
\end{tabular}
\end{center}
\end{table*}

\begin{table*}
\tiny \tiny \caption{\small{\label{tab:5} Constraints on the
temperature during reheating in the tachyon model with
superpotentail, obtained from Planck2018 TT, TE,
EE+lowE+lensing+BK14+BAO joint data at $68\%$ CL.}}
\begin{center}
\begin{tabular}{cccccc}
\\ \hline \hline \\ $\lambda$&$\alpha_{2}$& $\omega=-1$& $\omega=-\frac{1}{3}$
&$\omega=0$
\\
\hline
\\$0.1$&$1.41\times10^{-2}<\alpha_{2}<1.83\times10^{-2}$&$14.6\leq \log_{10}\left(\frac{T_{rh}}{GeV}\right)$
&$13.3\leq \log_{10}\left(\frac{T_{rh}}{GeV}\right)$&$9.91\leq \log_{10}\left(\frac{T_{rh}}{GeV}\right)$\\ \\
\hline
\\$0.4$&$0.68\times 10^{-2}<\alpha_{2}<0.91\times
10^{-2}$&$14.3\leq \log_{10}\left(\frac{T_{rh}}{GeV}\right)\leq
14.9$&$12.2 \leq \log_{10}\left(\frac{T_{rh}}{GeV}\right)
\leq 14.6$&$8.69 \leq \log_{10}\left(\frac{T_{rh}}{GeV}\right)\leq 14.2$\\ \\
\hline
\\$0.7$&$0.48\times 10^{-2}<\alpha_{2}<0.68\times 10^{-2}$&$12.3 \leq \log_{10}\left(\frac{T_{rh}}{GeV}\right)\leq 14.5$&
$9.43\leq \log_{10}\left(\frac{T_{rh}}{GeV}\right)\leq 13.9$&
$4.03\leq \log_{10}\left(\frac{T_{rh}}{GeV}\right)\leq 13.1$\\ \\
\hline
\\$0.9$&$0.17\times 10^{-2}<\alpha_{2}<0.33\times 10^{-2}
$&$11.5\leq \log_{10}\left(\frac{T_{rh}}{GeV}\right)\leq 14.3$&$6.32
\leq \log_{10}\left(\frac{T_{rh}}{GeV}\right)\leq 12.9$&
$\log_{10}\left(\frac{T_{rh}}{GeV}\right)\leq 10.8$\\ \\
\hline \hline
\end{tabular}
\end{center}
\end{table*}

\newpage
\section{A Short Discussion on the Case with Quantum Effects}

In this section, we study the consequence of presenting the quantum
effects on the tachyon's dynamics in the early universe. To this
end, following Ref.~\cite{Noj03}, we include the quantum effect by
considering the conformal anomaly. By considering ${\cal{N}}$
scalar, ${\cal{N}}_{1}$ vector fields, ${\cal{N}}_{2}$ gravitons,
${\cal{N}}_{\frac{1}{2}}$ spinor and ${\cal{N}}_{hd}$
higher-derivative conformal scalars, we have the anomalous trace of
the stress tensor as~\cite{Duf97,Noj01,Noj03}
\begin{equation}\label{eq58}
{\cal{T}}=b\left(\Upsilon+\frac{2}{3}\Box R\right)+b'\,G\,,
\end{equation}
wehere
\begin{equation}\label{eq59}
b=\frac{{\cal{N}}+6{\cal{N}}_{\frac{1}{2}}+12{\cal{N}}_{1}+611{\cal{N}}_{2}-8{\cal{N}}_{hd}}{120(4\pi)^{2}}\,,
\end{equation}
and
\begin{equation}\label{eq60}
b'=-\frac{{\cal{N}}+11{\cal{N}}_{\frac{1}{2}}+62{\cal{N}}_{1}+1411{\cal{N}}_{2}-28{\cal{N}}_{hd}}{360(4\pi)^{2}}\,.
\end{equation}
If we consider a warped de-Sitter space-time with a scale factor as
$a=e^{\frac{t}{L}}$ ($L$ is the curvature radius of the space-time),
the energy density and pressure corresponding to the quantum effects
are given by (see Refs.~\cite{Noj01,Noj02})
\begin{equation}\label{eq61}
\rho_{q}=-p_{q}=-\frac{6b'}{L^{4}}\,.
\end{equation}
Therefore, by considering the quantum effects and the tachyon field,
the Friedmann equations (\ref{eq5}) and (\ref{eq6}) take the
following forms
\begin{eqnarray}
\label{eq62}
\frac{3}{L^{2}}=\frac{\kappa^{2}\,V(\phi)}{\sqrt{1-\lambda\,\dot{\phi}^{2}}}-\frac{6\kappa^{2}\,b'}{L^{4}}\,,
\end{eqnarray}
\begin{eqnarray}
\label{eq63}
\frac{3}{L^{2}}=\kappa^{2}\,V(\phi)\,\sqrt{1-\lambda\,\dot{\phi}^{2}}-\frac{6\kappa^{2}\,b'}{L^{4}}\,.
\end{eqnarray}
Equations (\ref{eq62}) and (\ref{eq63}) are satisfied if
$\dot{\phi}=0$ or $\phi=\phi_{0}=constant$. Also, from equation of
motion (\ref{eq7}) we find $V'=0$. This means that in the presence
of the quantum effects, the potential of the tachyon field becomes
constant. We name this constant potential as $V_{0}$. Now, from
equation (\ref{eq62}) we have
\begin{eqnarray}
\label{eq64} \frac{1}{L^{2}}=-{\frac {1}{{
4\,b'}\,{\kappa}^{2}}}\pm\sqrt {{\frac {V_{0}}{6b'}}+{\frac
{1}{16\,{b'}^{2}{\kappa}^{4}}}} \,.
\end{eqnarray}
Considering that $b'$ is constant, equation (\ref{eq64}) give us
constant Hubble parameter. The constant Hubble parameter means
$\epsilon=\eta=0$. Even we consider small potential (as the case in
Ref.~\cite{Noj03}) and consider the quantum effect as the running of
inflation, there is still a problem. With a constant Hubble
parameter, the inflation phase would never have a graceful exit.
This problem remains even we assume the tachyon/phantom filed or a
canonical scalar field. If we insist to account for the quantum
effect in the inflation epoch, we may consider models with two
scalar fields with quantum effects (as has been done in
Ref.~\cite{Noj03}) or maybe nonminimal coupling models.

\section{Summary}
In this paper, we have studied the tachyon inflation with the
superpotential as an inflationary potential. In this regard, without
including any supersymmetry, we have used a form of the potential
which is similar to the one in the supergravity. By studying the
inflation in this model, we have obtained the slow-roll parameters
in terms of the superpotential. We have also explored the
perturbations in this model at both the linear and non-linear
levels. At the linear level, we have obtained the important
perturbation parameters, such as the scalar spectral index, the
tensor spectral index and the tensor-to-scalar ratio. At the
non-linear level, we have obtained a parameter named non-linearity
parameter which gives the amplitude of the non-gaussianity.
Considering that the amplitude of the non-gaussianity in the tachyon
model has a peak at the equilateral configuration, we have obtained
the non-linearity parameter in this configuration.

After that, we have sought for the observational viability of the
tachyon model with superpotential. In this regard, we have focused
on the model's parameters $\lambda$ and $\alpha_{2}$. By using the
observational constraints on the perturbations parameters, obtained
from Planck2018 TT, TE, EE+lowE+lensing+BAO+BK14 and Planck2018 TT,
TE, EE +lowE+lensing+BK14+BAO+LIGO and Virgo2016 data at $68\%$ CL,
we have obtained the observationally viable regions of the
parameters $\lambda$ and $\alpha_{2}$. We have also studied the
behavior of $r-n_{s}$ and $r-n_{T}$, in the background of Planck2018
TT, TE, EE+lowE+lensing+BAO+BK14 and Planck2018 TT, TE, EE
+lowE+lensing+BK14+BAO+LIGO and Virgo2016 data, respectively, at
$68\%$ CL and $95\%$ CL. Also, by using the observational
constraints on the scalar spectral index at $68\%$ CL and $95\%$ CL,
we have plotted the observationally viable ranges of the sound speed
and $\lambda$ which fulfill the constraints. By these numerical
analyses, we have obtained some constraints on the parameter
$\alpha_{2}$ for some sample values of $\lambda$ as
$\lambda=0.1,0.4,0.7$ and $0.9$. Our numerical analysis on the
perturbations parameters shows that, depending on the values of
$\lambda$, the tachyon model with superpotential is observationally
viable if $0.040\times10^{-2}\leq\alpha_{2}<0.1$ at $68\%$ CL and
$0.021\times10^{-2}\leq\alpha_{2}<0.1$ at $95\%$ CL. We have also
studied the non-gaussian feature of the primordial perturbations
numerically. By plotting the phase space of the parameters $\lambda$
and $\alpha_{2}$, we have presented the prediction of the tachyon
model with superpotential for the amplitude of the equilateral
non-gaussianity. We have shown that, for $\lambda>0.9$, to the
amplitude of the primordial non-gaussianity be consistent with
Planck2018 TTT, EEE, TTE and EET data, there are some constraints on
the parameter $\alpha_{2}$. Note that, for $\lambda\leq 0.9$, all
values of $\alpha_{2}$ lead to the observationally viable values of
$f_{NL}^{equil}$.

Another issue that has been studied in this paper, is the reheating
process after inflation by assuming the superpotential. To study
this process, we have found the e-folds number and temperature
during the reheating era in terms of the superpotential. Then, given
the relation between these parameters and scalar spectral, by using
the observational constraints on $n_{s}$, we have studied the
parameters $N_{rh}$ and $T_{rh}$ numerically. We have shown that, by
increasing the values of the e-folds number during the reheating
phase, the effective equation of state in the tachyon model with
superpotential increases from $-1$ to $\frac{1}{3}$ (corresponding
to the radiation dominated universe). We have also, studied the
behavior of the parameters $N_{rh}$ and $T_{rh}$ versus $n_{s}$ for
some sample values of $\lambda$ and $\alpha_{2}$. Our numerical
analysis shows that, in the tachyon model with superpotential, the
reheating is instantaneous for $\lambda<0.2$ and
$\alpha_{2}<1.53\times 10^{-2}$. In other cases, the reheating
process needs a few e-folds number to be completed. We have also
shortly discussed the case with presenting the quantum effects. It
seems that if we consider the models with single scalar fields, the
presence of quantum corrections makes the inflation phase permanent.
To have a graceful exit from inflation, we may consider the quantum
effects in two-field or nonminimal models.

{\bf Acknowledgement}\\
I thank the referee for the very insightful comments that have
improved the quality of the paper considerably.\\

\end{document}